\pdfoutput=1

\documentclass[12pt]{article}
\usepackage[doublespacing]{setspace}
\usepackage[margin = 1in]{geometry}
\usepackage{amsmath, amsthm, amssymb}
\usepackage{float}
\usepackage[titletoc,page]{appendix}

\usepackage{tabularx}
\newcolumntype{L}{>{\raggedright\arraybackslash}X}
\usepackage{csquotes}
\usepackage[authordate, giveninits=true, backend=biber, maxcitenames=3, maxbibnames=99, sortcites=true]{biblatex-chicago}
\usepackage[colorlinks=true, urlcolor=blue, citecolor=blue, filecolor=blue]{hyperref}
\usepackage{booktabs}
\usepackage{threeparttable}
\usepackage{caption}
\usepackage{graphicx}
\usepackage{enumitem}
\usepackage{titlesec}
\usepackage{cochineal}
\usepackage{mdframed}
\usepackage{footmisc}
\usepackage{tikz}
\usetikzlibrary{arrows.meta}
\titleformat{\section}{\normalfont\fontsize{16}{15}\bfseries}{\thesection}{1em}{}

\newtheorem{proposition}{Proposition}
\newtheorem{corollary}{Corollary}

\author{Andrew S. Rosenberg\thanks{Associate Professor, Department of Political Science, University of Florida. Email: \href{mailto:andrewrosenberg@ufl.edu}{andrewrosenberg@ufl.edu}. An R package implementing the diagnostic, \texttt{ferobust}, is available at \url{https://github.com/asrosenberg/ferobust}.}}
\date{\large This Version: \today\\[0.5em] \normalsize Word count: 6,101}
\title{\textsc{Reliable Panel Regression}\\[0.3em]
\large A Default Workflow for Slow-Moving, Mismeasured Variables}
\begin{document}
\maketitle
\setcounter{secnumdepth}{-2}

\begin{abstract}
\begin{singlespace}
\noindent Political scientists often interpret coefficient shrinkage under fixed effects as evidence that pooled associations are confounded. This paper shows why that inference is unreliable for slow moving, mismeasured regressors. Fixed effects can remove much of the signal and identify coefficients from within unit variation that is disproportionately measurement error, attenuating estimates toward zero. A lone fixed effects coefficient may therefore be unable to distinguish confounding from measurement error. I show that the attenuation depends on a regressor's empirical intraclass correlation and measurement reliability. I then propose a default workflow for panel regression. Researchers estimate reliability when possible, report pooled and fixed effects estimates with corrected within reliability, use partial identification bounds when the estimates share a sign, and report fixed effects as a within unit estimate when they do not. For variables with no reliability estimate, I introduce an autocorrelation frontier that bounds the attenuation factor directly. I conclude by applying this workflow to several published results to show that the data often cannot distinguish attenuation from confounding, and the workflow makes clear which case the researcher faces.

\medskip
\noindent\textbf{Keywords:} measurement error; fixed effects; panel data; partial identification; attenuation bias
\end{singlespace}
\end{abstract}
\thispagestyle{empty}
\newpage
\setcounter{page}{1}

\section*{Introduction}

Does democracy cause economic growth? Is there a resource curse? Do stronger bureaucracies protect democracy? These are important questions in comparative politics and international relations and scholars typically use one-way or two-way fixed effects models to answer them. Often, a large, significant pooled relationship shrinks once one adds fixed effects. Researchers usually interpret that shrinkage as evidence that the former association was confounded. One then treats the latter estimate as the more credible result because it controls for time-invariant confounders and common shocks.

This paper shows why that interpretation is often wrong. Fixed effects remove between-unit variation and use within unit change to identify coefficients. That is useful when the within unit change is informative, but many variables political scientists care about, including regime type, bureaucratic quality, military capability, trade dependence, national income, and oil rents, move slowly and are measured with error. For these variables, much of the important variation lies between units or in long-run trends. The short run movement left after fixed effects can contain significant measurement error. In that case, fixed effects do not simply remove confounding. They also make the regressor less reliable, so the estimated coefficient is pulled toward zero even when the underlying relationship exists.

As a result, finding shrinkage in fixed effects models is ambiguous. It may mean that the pooled estimate was biased, but it may also mean that fixed effects have removed much of the signal and left the estimate to be identified from low reliability within unit variation. The usual interpretation assumes the first story and can push researchers toward substantive null conclusions: democracy does not cause growth, oil does not weaken accountability, bureaucratic quality does not matter. The more cautious conclusion is that the fixed effects estimate may not identify the effect with the available measurement information.

In this paper, I propose a reporting standard for this problem. The main diagnostic is the corrected within reliability of the regressor, denoted $\lambda_w$. This quantity tells the researcher how much of the variation used by fixed effects is signal rather than measurement error. To calculate it, the researcher needs the empirical intraclass correlation (ICC) of the observed regressor and an estimate or range of estimates for the regressor's measurement reliability. The empirical ICC measures how much of a variable's variation lies between units. Slow moving variables have high ICCs, and fixed effects discard exactly that variation. The reliability term captures how precisely a persistent variable is measured.

I argue that researchers should report pooled OLS, fixed effects, the empirical ICC of the key regressor, and the corrected within reliability. A simple sign test then determines what should be reported. If the estimates share a sign and fixed effects is smaller in magnitude, researchers should report partial identification bounds on the effect. If the estimates have opposite signs, they should report fixed effects as a within unit result. If fixed effects is larger in magnitude, the result is a complement rather than an attenuation problem. For variables with no reliability estimate, I introduce an autocorrelation frontier that asks whether measurement error can plausibly explain the shrinkage using the regressor's own persistence. The Diagnostic section develops each case. I provide an open-source R package, \texttt{ferobust} \autocite{ferobust2026}, that automates the diagnostic.

I apply this reporting standard to the case of the effect of bureaucratic quality on economic growth \autocite{cornell2020}. In this example, the fixed effects estimate is positive but insignificant. Once measurement error is incorporated, however, the identified set excludes zero under a broad range of reliability. The fixed effects null therefore cannot be interpreted as evidence of no within unit effect.

A broader audit of the literature shows that, in most cases, the reporting standard does not overturn a null finding. It either tells researchers to report fixed effects because the pooled and within unit estimates have opposite signs, shows that measurement error is unlikely to explain a null finding, or concludes that the data cannot distinguish attenuation from confounding. The point of this article is not to argue that fixed effects models are usually wrong. Rather, the point is that a fixed effects coefficient by itself does not tell scholars which of these situations they are in. Accordingly, this paper is similar to other recent work that implores quantitative political scientists to be clear-eyed and humble when drawing conclusions from their work \autocite[e.g.,][]{arel2026quantitative}.

The argument also complements the literature on pathologies in two-way fixed effects designs with staggered binary treatments and heterogeneous effects \autocite{goodmanbacon2021,dechaisemartin2020,callaway2021,sunandabraham2021}. That literature addresses difference-in-differences designs with binary treatments. But many panel studies in political science instead use continuous, slow-moving regressors measured across countries or organizations over time, which it does not cover \autocite[247]{chiuetal2026}. For the modal panel regression in comparative politics and international relations, the danger is measurement error amplified by fixed effects, not negative weighting.

\section{The Hidden Cost of Fixed Effects}\label{sec:hiddencost}

\subsection{Measurement Error Amplification}

To see this issue, consider Polity scores. A country's Polity score is usually relatively constant over time, but there is typically a lot of cross-sectional variation. For instance, Saudi Arabia and Norway have very different scores, and most of the variable's overall signal is in those cross-country differences. Polity also includes measurement error because coders disagree, the historical record is thin, and later revisions rewrite earlier country-year values.

Country fixed effects subtract a country's average Polity score from its score in a given year, so the model uses within-country variation around its average to identify coefficients. These models are useful under time-invariant confounding. However, for a slow moving variable like Polity, the country average is not only confounding, it is also where much of the real variation is. After country fixed effects are added, the model relies on the remaining movement within countries. Some of that movement reflects real regime change. Some of it reflects coding disagreement, later revisions, or small annual changes with little substantive meaning. Fixed effects can therefore leave the researcher with a noisier version of the original variable.

The formal result follows from a standard panel model:
\[
y_{it} = \alpha_i + \beta x_{it}^* + \varepsilon_{it},
\]
where $y_{it}$ is the outcome for unit $i$ in period $t$, $\alpha_i$ is a unit fixed effect, $x_{it}^*$ is the true regressor, and $\varepsilon_{it}$ is the error term. The researcher observes a noisy version,
\[
x_{it} = x_{it}^* + u_{it},
\]
where $u_{it}$ is classical measurement error: zero mean, uncorrelated with the true regressor and the outcome error. The overall \emph{reliability} of the observed regressor is
\[
\lambda = \frac{\operatorname{Var}(x^*)}{\operatorname{Var}(x)},
\]
the share of observed variance that is signal rather than noise.

With no fixed effects, classical measurement error attenuates the coefficient toward zero by the factor $\lambda$. With fixed effects, the quantity that matters is not the overall reliability but the reliability of the within unit variation in $x_{it}$. Call it the within reliability $\lambda_w$, and, importantly, the fixed effects estimator converges to $\beta \lambda_w$, not $\beta$. How much reliability survives the within transformation depends on how much of the regressor's variance lies between units. Let $\operatorname{ICC}^*$ be the intraclass correlation of the true regressor,
\[
\operatorname{ICC}^* = \frac{\operatorname{Var}(\bar{x}_i^*)}{\operatorname{Var}(x^*)},
\]
the share of variance that lies between rather than within units. A high-ICC variable barely moves within units over time, and \textcite{griliches1986} show that, under classical measurement error,
\[
\lambda_w =
\frac{(1-\operatorname{ICC}^*)\lambda}{1-\operatorname{ICC}^*\lambda}.
\]
The more persistent the true regressor, the higher $\operatorname{ICC}^*$ and the lower the within reliability.

One cannot use this formula in practice because $x_{it}^*$ is unobserved and the empirical ICC is itself attenuated by measurement error. Under classical measurement error with serially uncorrelated noise and large $T$,
\[
\widehat{\operatorname{ICC}} = \operatorname{ICC}^* \lambda.
\]
Substituting $\operatorname{ICC}^* = \widehat{\operatorname{ICC}}/\lambda$ into the Griliches-Hausman expression yields a corrected within reliability formula:
\begin{equation}\label{eq:within_rel}
\lambda_w =
\frac{\lambda - \widehat{\operatorname{ICC}}}{1 - \widehat{\operatorname{ICC}}}.
\end{equation}

One needs two quantities for this computation: the empirical ICC of the observed regressor, computed directly from the data, and the overall reliability $\lambda$, from a measurement model, a second measure, published reliability evidence, or a defended range.

Equation~\ref{eq:within_rel} is also a diagnostic. A slow moving regressor with a high empirical ICC can have high overall reliability, yet low within reliability. If $\widehat{\operatorname{ICC}} = 0.75$ and $\lambda = 0.90$, then
\[
\lambda_w = \frac{0.90 - 0.75}{1 - 0.75} = 0.60,
\]
so a fixed effects model recovers only about 60 percent of the coefficient.

\begin{figure}[!htbp]
\centering
\includegraphics[width=\textwidth]{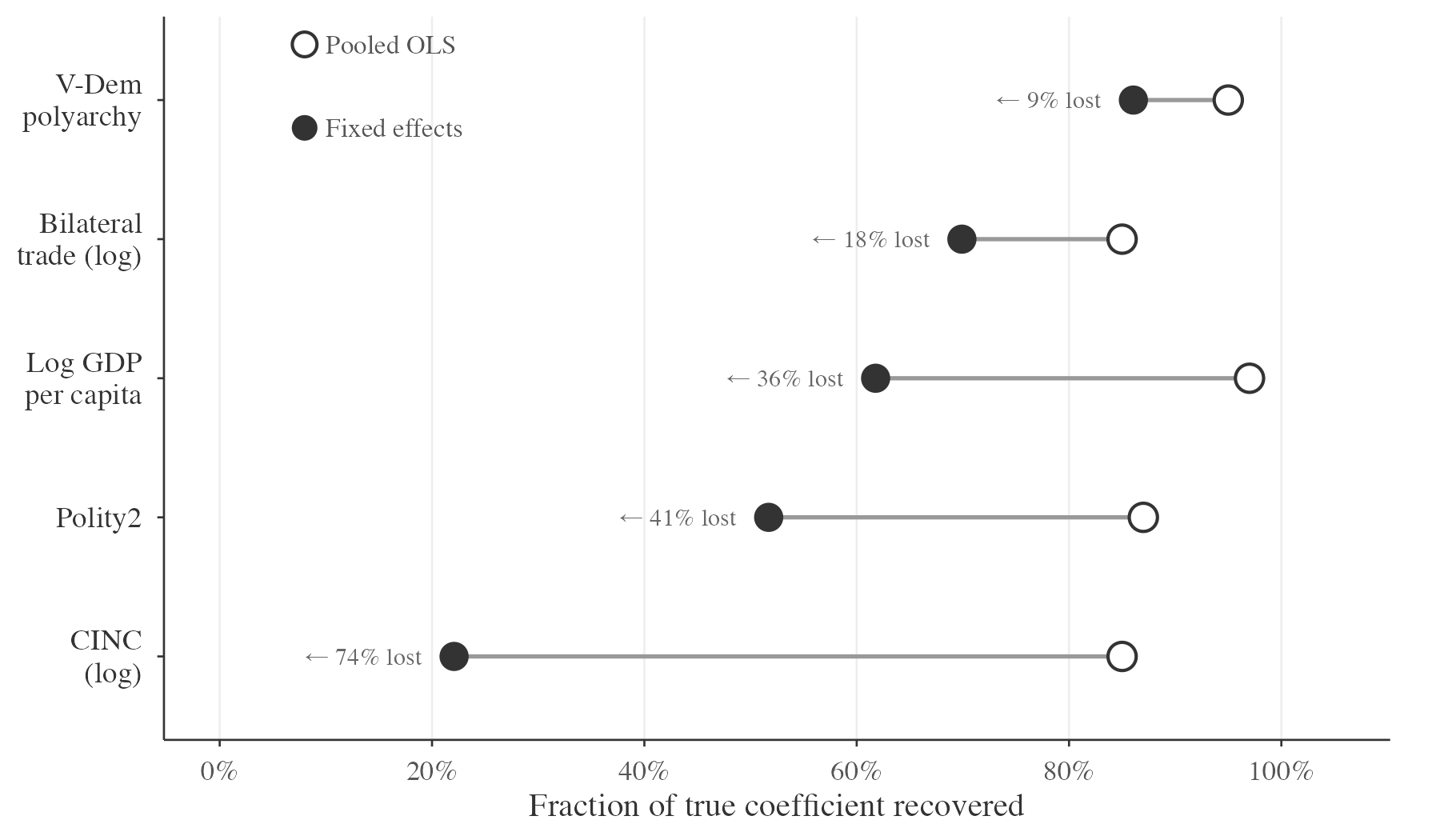}
\caption{The hidden cost of fixed effects. Open circles show coefficient recovery under pooled OLS, equal to the overall reliability $\lambda$; filled circles show recovery under fixed effects, equal to the corrected within reliability $\lambda_w$ of Equation~\ref{eq:within_rel}. Each variable is plotted at an illustrative reliability within its defensible range.}
\label{fig:hidden_cost}
\end{figure}

Figure~\ref{fig:hidden_cost} shows the attenuation problem for five common variables in political science. Overall reliability is high for all five, so none would look problematic in a pooled regression. Once the model uses only within unit change, though, Polity and log GDP per capita lose roughly half their signal, and CINC loses nearly three quarters. Thus, a measure can be reliable in levels yet unreliable in the variation fixed effects use.

This derivation assumes serially uncorrelated measurement error, which leaves most of the noise in the within dimension and makes the diagnosis conservative. I take up persistent, coder- or source-specific error below.

\subsection{The Scope of the Problem}\label{sec:scope}

\begin{figure}[!htbp]
\centering
\includegraphics[width=\textwidth]{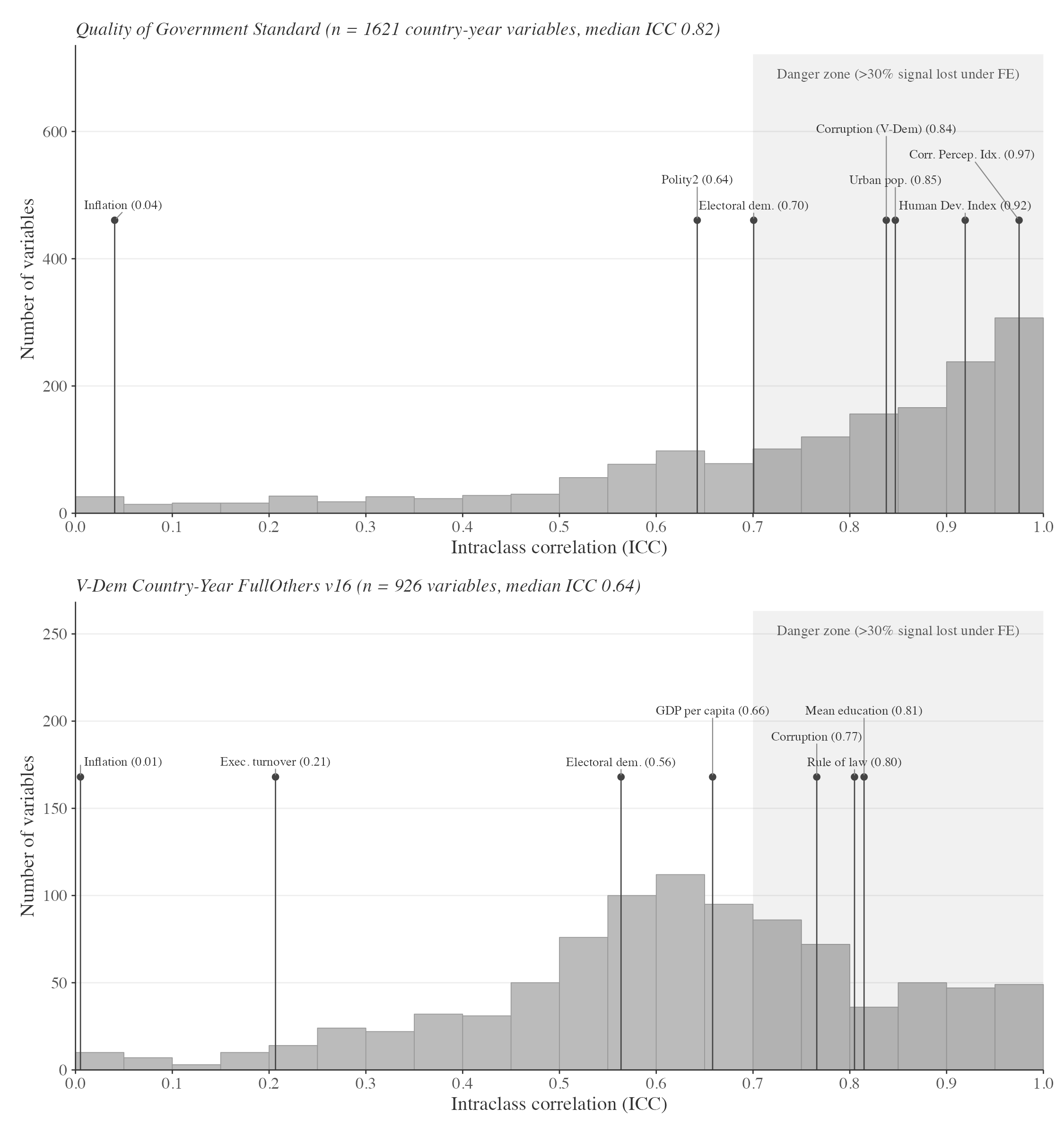}
\caption{Distribution of empirical ICCs, $\widehat{\operatorname{ICC}} = \operatorname{Var}(\bar{x}_i)/\operatorname{Var}(x)$, across two major political science data sources. Top: Quality of Government Standard, with 1,621 country-year variables. Bottom: V-Dem Country-Year v16, with 926 variables. Shaded regions mark the danger zone where fixed effects destroy more than 30\% of the signal.}
\label{fig:icc_hist}
\end{figure}

Are these examples unusual? Many panel variables in political science move slowly, and high persistence is exactly when fixed effects amplify measurement error. Figure~\ref{fig:icc_hist} audits every numeric country-year variable in the Quality of Government and V-Dem Country-Year datasets \autocite{qog2025, vdem2026}, and it computes the empirical ICC\footnote{$\widehat{\operatorname{ICC}} = \operatorname{Var}(\bar{x}_i)/\operatorname{Var}(x)$} for each.

The results show that high persistence is widespread in both datasets. In the Quality of Government data, 67 percent of the variables have ICCs above 0.70 and 34 percent exceed 0.90. V-Dem is less extreme but still contains a significant percentage of highly persistent variables: 37 percent exceed 0.70 and 10 percent exceed 0.90, and many of its core institutional and regime measures sit in the ``danger zone.''

This exploratory analysis shows the enormous potential for fixed effects attenuation in the political science literature, but not that any particular variable is severely attenuated, because severity depends on both the empirical ICC \emph{and} reliability. A high-ICC variable with near-perfect reliability may retain meaningful within signal, while one with moderate reliability may not. Appendix~\ref{app:reliability} reports estimated within-reliability for 20 commonly used variables. I draw on \textcite{pemstein2018} for V-Dem indices and the relevant measurement literature otherwise.\footnote{Other sources include \textcite{treier2008}, \textcite{johnson2013}, \textcite{feenstra2015}, and \textcite{kaufmann2011}.} Nine of the 20 variables lose more than 30 percent of their coefficient under unit fixed effects, and two of them, the Corruption Perceptions Index and the trade-to-GDP ratio, lose more than 80 percent.

Two-way fixed effects make the problem worse because year fixed effects also remove common time variation, which for variables such as income or capability is often substantively meaningful. The two-way FE diagnostic replaces the unit ICC with the share of variance absorbed jointly by unit and year:
\begin{equation}\label{eq:twfe}
\lambda_w^{\text{2FE}} =
\frac{\lambda - \widehat{\operatorname{ICC}}_{uy}}{1 - \widehat{\operatorname{ICC}}_{uy}}.
\end{equation}
In a balanced panel with orthogonal unit and year components, $\widehat{\operatorname{ICC}}_{uy} = \widehat{\operatorname{ICC}}_u + \widehat{\operatorname{ICC}}_t$. In unbalanced panels, I compute the joint absorbed component directly. The difference between one-way and two-way fixed effects can be large. In the \textcite{treisman2015} democracy and growth data, log GDP per capita has $\widehat{\operatorname{ICC}}_u = 0.70$ and $\widehat{\operatorname{ICC}}_t = 0.26$ with $\lambda = 0.97$. In this case, one-way country fixed effects imply $\lambda_w = 0.90$, while two-way fixed effects imply $\lambda_w^{\text{2FE}} = 0.25$, which leaves only one quarter of the coefficient recoverable under the measurement error model. These calculations identify when an estimate is likely to rely on low reliability within variation. They do not show that every high ICC variable is problematic, and the reporting standard below evaluates reliability on a case-by-case basis.

\subsection{Precisely Wrong: The Inference Problem}

\begin{figure}[!htbp]
\centering
\includegraphics[width=0.85\textwidth]{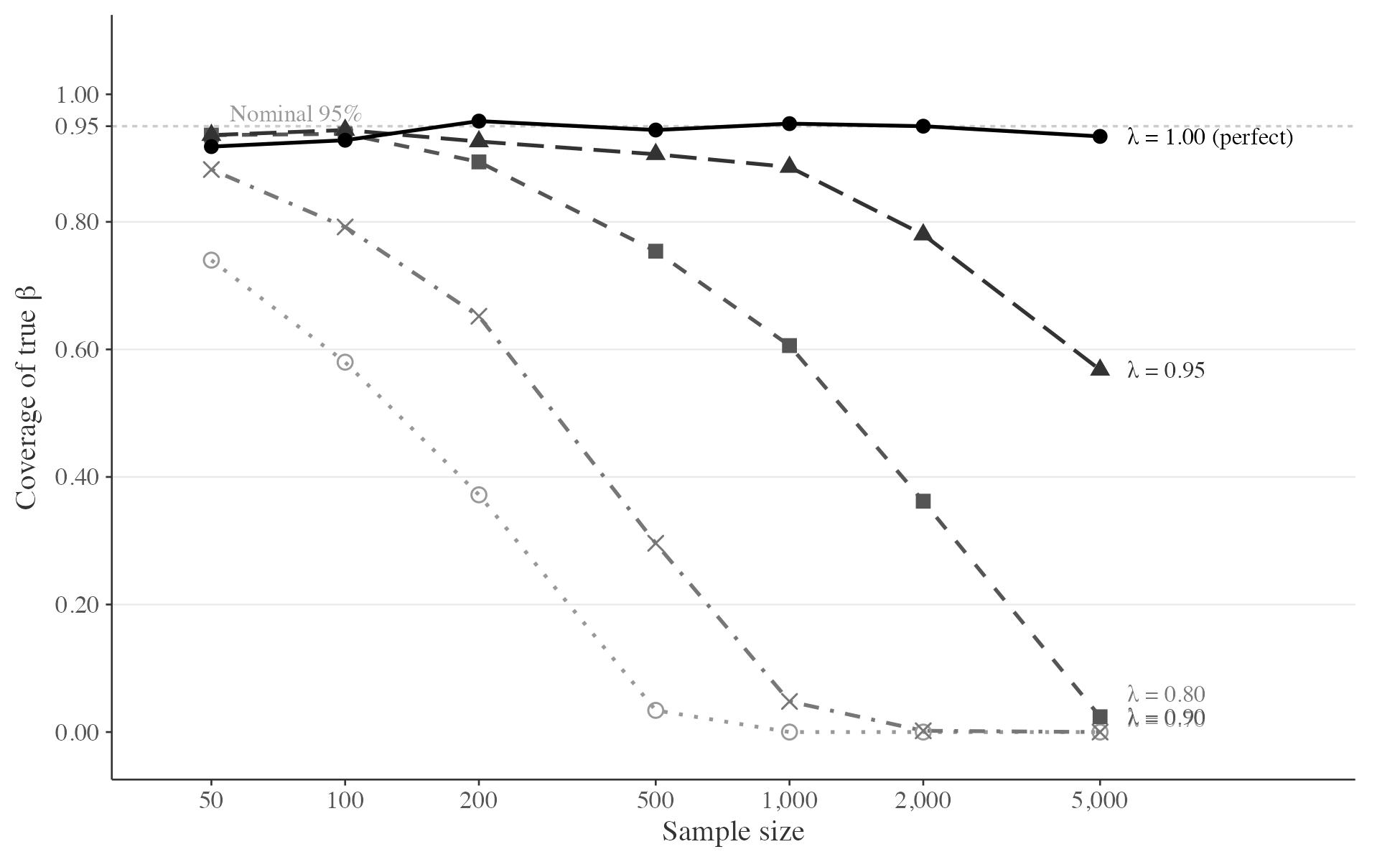}
\caption{Coverage of nominal 95\% confidence intervals for $\beta = 0.5$, by sample size and reliability. With perfect measurement, $\lambda = 1$, coverage holds at 95\%. With measurement error, larger samples produce tighter intervals centered on the attenuated estimate, $\beta\lambda$, and coverage of the true $\beta$ converges to zero.}
\label{fig:precisely_wrong}
\end{figure}

Measurement error also changes how researchers should interpret the statistical significance of coefficients.Scholars typically assume large samples and small standard errors bring estimates closer to the ``truth,'' but under fixed effects the opposite can hold. The sampling distribution of the FE estimate is centered on $\beta\lambda_w$, not $\beta$, and the bias $\beta(1-\lambda_w)$ does not shrink as sample size increases. The sampling variance shrinks while the estimator remains centered on the attenuated value, so the confidence interval tightens around the wrong value. Counterintuitively, a small panel's interval may cover the true effect, while a large panel's tighter interval is significant, yet excludes it. Figure~\ref{fig:precisely_wrong} demonstrates this issue. When reliability is perfect, coverage stays near the 95 percent level as $N$ increases. But with even modest attenuation, larger samples make the problem worse. The coverage of the true value remains close to nominal for $\lambda = 0.95$ through moderate sample sizes before falling sharply, deteriorates much earlier for $\lambda = 0.90$, and collapses almost immediately for lower values of $\lambda$. Put simply, a ``significant'' finding may mean that one's dataset is large enough to estimate the attenuated quantity, $\beta\lambda_w$, with great precision while excluding the true effect, $\beta$. Appendix~\ref{app:extended_sim} confirms this in a 243-cell simulation, where bare fixed-effects intervals cover the truth in 4.8 percent of replications.

\section{The Diagnostic}\label{sec:diagnostic}

Which estimate should a researcher report? The diagnostic answers this using only quantities the researcher already has: the pooled OLS coefficient, the fixed-effects coefficient, the empirical ICC of the key regressor, and, when available, a defensible reliability estimate. It proceeds in two steps. The researcher first settles the two measurement inputs, the ICC and a reliability estimate, then applies a sign test that determines which result to report.

The researcher estimates reliability from the strongest available source. For measures that report observation-level uncertainty, such as the V-Dem indices, reliability follows from the posterior uncertainty. When two independent measures of the same construct exist, as for democracy, their association supplies external information about reliability. For common variables with no attached uncertainty, published reliability estimates serve when the literature provides them. When none of these sources exists, I describe a frontier of reliability values for scholars to use.

Why not estimate the noise variance instead of bounding it? Under transitory error, contrasting estimators that difference the data at different lengths point-identifies reliability \autocite{griliches1986, meijerspierdijkwansbeek2017, wansbeek2001}. This strategy fails twice over for the variables in Figure~\ref{fig:icc_hist}. Coder-based error is plausibly persistent, and persistent error attenuates short and long differences alike, so the contrast carries little information, and error as persistent as the signal carries none. This is the non-identification behind Proposition 2 below. Even transitory error leaves the contrast weakly identified, because the differences of a slow moving regressor are tiny: $\mathrm{Var}(\Delta_k x^*) = 2(1-\phi^k)\,\mathrm{Var}(x^*)$ with $\phi$ near one. Lag instruments inherit the same weakness near a unit root \autocite{blundell1998}. When a long panel and transitory error can be defended, Griliches-Hausman estimation simply supplies the $\lambda$ that the bounds take as input. Otherwise the frontier states the persistence assumption rather than fixing it at zero.

For the V-Dem indices I compute reliability from posterior uncertainty but report wider defended ranges, because posterior standard deviations capture coder disagreement and model uncertainty, not scale validity or systematic coder bias. The same logic answers a natural alternative: propagating the posterior draws of the index through the regression \autocite[e.g.,][]{treier2008, pemstein2018}. Every draw contains the measurement noise, so the fixed effects estimate in every draw is centered on $\beta\lambda_w$, and averaging over draws widens the interval around the wrong point estimate. The workflow instead uses the posterior to estimate the noise share and correct for it. Draw propagation also requires a measurement model, which most variables in Figure~\ref{fig:icc_hist} lack. Appendix~\ref{app:reliability} gives the details.

For a variable with no attached uncertainty and no alternative measure, reliability is not identified, and the relevant object is the within-reliability frontier of Proposition 2, which bounds $\lambda_w$ directly under an assumption about error persistence.

The empirical ICC places the application on the attenuation scale of Equation~\ref{eq:within_rel}. At an ICC near 0.70 and reliability near 0.85, the corrected within-reliability falls to roughly 0.50, so fixed effects recover only about half the coefficient under the measurement-error model. The ICC tells the researcher how severe attenuation may be once reliability is specified.

With both inputs in hand, the sign test determines what kind of result to report. When the pooled OLS and fixed effects estimates share a sign and the latter is smaller in magnitude, the researcher uses the defensible range of reliability values to report an \emph{identified set}. When the fixed effects estimate is larger in magnitude than pooled OLS, pooled OLS no longer bounds $|\beta|$ from above. This complement case supports the within-unit result, reported as a one-sided floor on $|\beta|$ rather than as an identified set. When the two estimates have opposite signs, the bounds do not apply, and the fixed effects estimate should be reported as a within-unit estimate.

The diagnostic changes how one interprets when coefficients shrink under fixed effects. Same-sign shrinkage does not by itself show that the pooled estimate was confounded away. It may instead show that the fixed effects estimate depends on low-reliability within-unit variation. When the resulting identified set includes both zero and substantively meaningful effects, one should not conclude that the ``true'' effect is zero, but rather that the available design and data cannot distinguish a null from a meaningful effect. Appendix~\ref{app:diag_accuracy} characterizes the rule's operating characteristics.

\section{Partial Identification}\label{sec:bounds}

\subsection{The identified set for the coefficient}

The diagnostic tells the researcher when to report bounds rather than a point estimate. The two-sided identified set applies when pooled OLS and fixed effects estimates have the same sign and fixed effects is smaller in magnitude. The question is how much of the difference between the two effects reflects the fixed effects estimator removing time-invariant confounding and how much reflects measurement error attenuation. The answer is generally not point identified. When one can assume that time-invariant confounders bias pooled OLS away from zero, then pooled OLS provides an upper bound while the de-attenuated fixed effects estimate provides the lower bound. Narrow bounds mean the within-unit variation is informative even after correction, while wider bounds mean the design cannot separate attenuation from confounding. When fixed effects is larger than pooled OLS, one should instead report a one-sided floor on $|\beta|$.

\begin{proposition}[Partial Identification]\label{prop:bounds}
Suppose three conditions hold. First, pooled OLS is biased away from zero in probability limit, so that
\[
|\mathrm{plim}\,\hat{\beta}_P| \geq |\beta|.
\]
Second, classical measurement error attenuates the fixed effects estimate toward zero, so that
\[
\mathrm{plim}\,\hat{\beta}_{FE} = \beta\lambda_w,
\]
with $\lambda_w \in (0,1]$. Third, the true overall reliability of the regressor lies in the defended interval $[\lambda_{\min},\lambda_{\max}]$. Then a conservative outer identified set for $\beta$ is
\[
\mathcal{B}
=
\bigcup_{\lambda \in [\lambda_{\min},\lambda_{\max}]}
\left[
\min\!\left\{
\frac{\hat{\beta}_{FE}}{\lambda_w(\lambda,\widehat{\operatorname{ICC}})},
\hat{\beta}_P
\right\},
\max\!\left\{
\frac{\hat{\beta}_{FE}}{\lambda_w(\lambda,\widehat{\operatorname{ICC}})},
\hat{\beta}_P
\right\}
\right],
\]
where $\lambda_w(\lambda,\widehat{\operatorname{ICC}})$ is given by Equation~\ref{eq:within_rel}.
\end{proposition}

Each value of $\lambda$ in the range gives one de-attenuated fixed effects coefficient; pairing it with pooled OLS and taking the union over the range produces $\mathcal{B}$. The first condition is the main assumption required. Pooled OLS bounds $|\beta|$ from above only when omitted variables bias it away from zero. In comparative politics and IR this assumption usually holds, since the leading omitted variables push the pooled estimate the same way as the hypothesized effect. However, when reverse causation or omitted variables could instead pull pooled OLS toward zero, the pooled estimate no longer bounds $|\beta|$, and the researcher reports the relaxed one-sided bound of Appendix~\ref{app:bounds}.

\begin{corollary}[Sign-agreement rule]\label{cor:sign}
Under the conditions in Proposition~\ref{prop:bounds}, pooled OLS and fixed effects share the sign of $\beta$ in probability limit. If the pooled OLS and fixed effects estimates have different signs in the sample, the bounds in Proposition~\ref{prop:bounds} do not apply.
\end{corollary}

Classical measurement error attenuates the estimate toward zero without reversing the sign. Signed confounding pushes pooled OLS farther from zero in the direction of $\beta$. Finding opposite signs in pooled and fixed effects estimates is therefore evidence that the between-unit and within-unit associations differ, and the fixed effects estimate should be reported as a within-unit estimate.

The width of $\mathcal{B}$ is itself a diagnostic. A wide identified set means that the data cannot sharply separate attenuation from confounding, so a single fixed effects coefficient creates false precision or confidence in one's conclusions.

The \textcite{imbens2004} confidence interval provides inference for the identified set. It propagates sampling error in the two estimates while treating the ICC and reliability range as fixed inputs. Simulations find this plug-in conservative (Appendix~\ref{app:diag_accuracy}).

\subsection{When no measure of reliability is available}\label{sec:frontier}

The identified set depends on the within reliability $\lambda_w$. When the researcher can defend an overall reliability $\lambda$, Equation~\ref{eq:within_rel} maps that value and the empirical ICC into $\lambda_w$. But some variables provide too little information to defend a value of $\lambda$. This is common when the researcher has one observed value per unit period, no measurement model, and no second measure of the same construct. Once measurement error may persist over time, the observed data cannot separate stability in the underlying construct from stability in the error. One additional assumption can still help. If the researcher places an upper bound on how persistent the measurement error can be, the autocorrelation of the observed variable gives a lower bound on $\lambda_w$: a series more persistent than the error is allowed to be must contain some real signal.

\begin{proposition}[Within reliability frontier]\label{prop:frontier}
Let $z_{it} = x_{it} - \bar{x}_i$ be the unit demeaned observed regressor. Write
\[
z_{it} = s_{it} + e_{it},
\]
where $s_{it} = x^*_{it} - \bar{x}^*_i$ is the unit demeaned signal and $e_{it} = u_{it} - \bar{u}_i$ is the unit demeaned measurement error. Let
\[
\lambda_w = \frac{\operatorname{Var}(s)}{\operatorname{Var}(z)}.
\]
Let $\rho_z(k)$, $\rho_s(k)$, and $\rho_e(k)$ be the lag $k$ autocorrelations of $z$, $s$, and $e$. Assume that the signal autocorrelation cannot exceed one, so $\rho_s(k) \leq 1$, and that the error autocorrelation is bounded above by $\psi_{\max}^{\,k}$, so $\rho_e(k) \leq \psi_{\max}^{\,k}$ for some $\psi_{\max} \in [0,1)$. Then
\begin{equation}\label{eq:frontier}
\lambda_w \geq
\max\left\{0,\; \max_{k \geq 1}
\frac{\rho_z(k) - \psi_{\max}^{\,k}}{1 - \psi_{\max}^{\,k}}
\right\}.
\end{equation}
The bound is computed from the autocorrelation function of the demeaned regressor and the single assumption $\psi_{\max}$. If $\psi_{\max} = 1$, the researcher allows measurement error to be as persistent as the signal, and the useful lower bound collapses to zero. (Proof in Appendix~\ref{app:proofs}.)
\end{proposition}

The parameter $\psi_{\max}$ is the researcher's assumption about how persistent measurement error can be. Setting $\psi_{\max}=0$ assumes that measurement error does not persist over time and gives the tightest lower bound. Larger values allow more persistent error, and $\psi_{\max}=1$ leaves no useful bound. The researcher should report the frontier across a few defensible values of $\psi_{\max}$. If the lower bound stays high even when persistent error is allowed, fixed effects retain meaningful signal. If the lower bound falls near zero, the fixed effects estimate may be severely attenuated. Appendix~\ref{app:frontier_val} shows that the frontier covers the true $\lambda_w$ when $\psi_{\max}$ is at least as large as the true error persistence.

\begin{corollary}[Certification]\label{cor:certify}
Let $\rho_z(k)$ be the lag-$k$ within unit autocorrelation of the demeaned observed regressor, let
\[
\underline{\lambda}_w(\psi_{\max})
= \max\!\Big\{0,\ \max_{k\ge 1}\tfrac{\rho_z(k)-\psi_{\max}^{\,k}}{1-\psi_{\max}^{\,k}}\Big\}
\]
be the frontier floor of Proposition~\ref{prop:frontier}, and let $r \equiv |\hat\beta_{FE}/\hat\beta_P|$ be the shrinkage ratio. Suppose pooled OLS and fixed effects share a sign with $|\hat\beta_{FE}| < |\hat\beta_P|$. Then the frontier rules out the measurement error reading of the shrinkage, in the sense that the de-attenuated fixed effects estimate $\hat\beta_{FE}/\lambda_w$ cannot reach $\hat\beta_P$ for any frontier-admissible $\lambda_w$, if and only if
\[
\underline{\lambda}_w(\psi_{\max}) > r .
\]
At $\psi_{\max}=0$ the condition is $\max_k \rho_z(k) > r$, for which $\rho_z(1) > r$ is sufficient. (Proof in Appendix~\ref{app:proofs}.)
\end{corollary}

The corollary is a simple test: compare the regressor's within unit autocorrelation to the shrinkage ratio. When the within variation is more persistent than the coefficient is shrunk, so $\rho_z(1) > r$, measurement error cannot explain the shrinkage, and the identified set lies strictly inside the cross-sectional association. The measurement error explanation requires the reverse, $\rho_z(1) \le r$: a regressor persistent in levels but transitory within units. That combination is uncommon, because slow moving regressors carry $\rho_z(1)$ near one and danger zone shrinkage makes $r$ small. The frontier therefore mainly asks whether shrinkage can be explained by measurement error, and usually shows it cannot, defending a within unit result rather than rescuing a null. The rescues this paper reports, such as \textcite{cornell2020} below, come instead from a resolved reliability paired with mild shrinkage.

\section{Applications}\label{sec:apps}

I work through the bureaucracy and growth study of \textcite{cornell2020} to demonstrate the workflow, and then conduct a larger audit of the literature. I used Claude Opus 4.8 (Anthropic, accessed via \texttt{claude.ai}) in May 2026 to assist with this audit. Specifically, I used Claude to 1. identify candidate published studies whose designs fit the diagnostic's scope conditions, 2. locate and download the corresponding replication datasets, and 3. draft R code that applies the \texttt{ferobust} package to each application. I used the model as provided, without fine-tuning or training on my own data. I independently re-ran and verified all reported estimates, diagnostics, and verdicts against the original studies, and the full analysis is reproducible from the replication code without any AI tool. I have no competing interest in Anthropic.

Table \ref{tab:audit} then summarizes the audit results. The main pattern is that the within unit effect is often not identified. A single fixed effects coefficient does not reveal whether the researcher is looking at confounding removed, measurement error attenuation, a sign flip, or a complement case. The rare case in which the correction overturns a fixed effects null is credible precisely because the workflow does not rescue most null findings.

\subsection{Does bureaucratic quality increase economic growth?}\label{sec:cornell}

\textcite{cornell2020} use fixed effects models to ask whether bureaucratic quality raises growth. The pooled association is large and positive, but the two-way FE estimate is small and insignificant, so they interpret the latter estimate as the more credible result. Replicating their specification, a regression of five year ahead growth on V-Dem's index of impartial administration with country and year fixed effects gives a coefficient of $0.153$ ($p = 0.158$). The estimate is positive, smaller than the pooled estimate of $0.184$ ($p = 0.002$), and indistinguishable from zero at conventional levels.

However, that fixed effects coefficient does not settle the question. It is consistent with no effect, but once measurement error is allowed, it is also consistent with an effect close to the pooled association. The sign test passes because pooled OLS and fixed effects are both positive. The issue is how much of the shrinkage reflects confounding and how much reflects attenuation. In this specification, country and year fixed effects jointly absorb $77.5$ percent of the variance in the V-Dem index, so $\widehat{\operatorname{ICC}} = 0.775$. Since V-Dem reports observation level uncertainty, I use the measurement information in the index rather than a default reliability value. At $\lambda = 0.85$, the corrected within reliability is $\hat\lambda_w = (0.85 - 0.775)/(1 - 0.775) = 0.33$. Correcting the fixed effects estimate across the defended reliability interval $[0.85, 0.95]$ gives a within country effect at or above the pooled association. The reported set is therefore the conservative one sided union $[0.184, 0.458]$ (Appendix~\ref{app:bounds}), and the Imbens-Manski 95\% interval is $[0.08, 1.01]$. Both estimates exclude zero.

\begin{figure}[!htbp]
\centering
\includegraphics[width=0.95\textwidth]{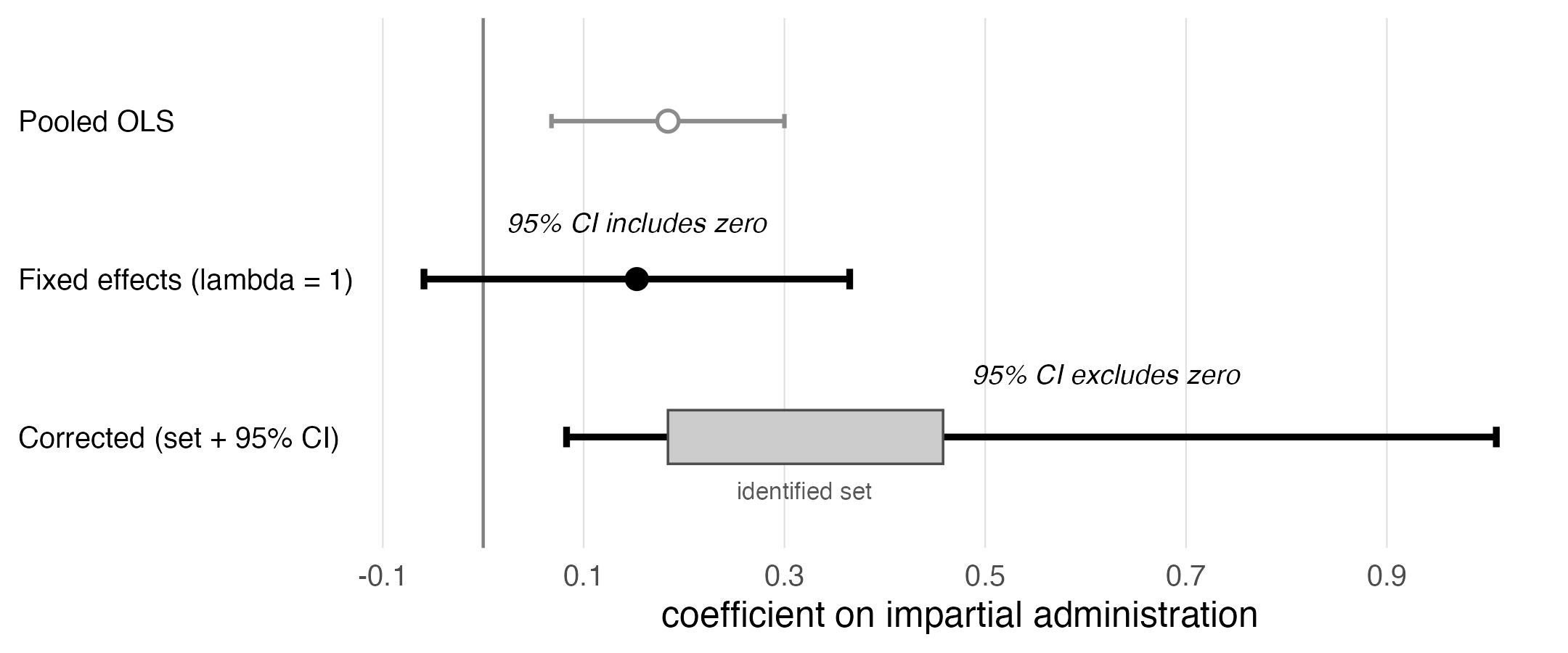}
\caption{Bureaucracy and growth re-analysis}
\label{fig:bounds_ckt}
\end{figure}

Table~\ref{tab:cornell} and Figure~\ref{fig:bounds_ckt} report the results. The ordinary fixed effects interval assumes perfect measurement and includes zero. The Imbens-Manski interval uses the defended reliability range and excludes zero. The correction does not make the fixed effects estimate significant. It changes the interpretation of the fixed effects null. If the index were perfectly reliable, the null would be evidence against a within country effect. But at reliability values supported by the measurement evidence, the same estimate is consistent with an attenuated effect as large as the pooled association. Reporting only the fixed effects coefficient therefore amounts to assuming $\lambda = 1$ for an index that is not measured perfectly. The result is also sensitive to nonclassical measurement error. If measurement/coding errors in the V-Dem index is sufficiently correlated with growth, the interpretation of the case becomes ambiguous. Correlated random effects does not eliminate the problem, because its within component inherits the same attenuation as fixed effects (Appendix~\ref{app:cre}).

\begin{table}[!htbp]
\centering
\begin{threeparttable}
\caption{Bureaucracy and Growth: Diagnostic on \textcite{cornell2020}}\label{tab:cornell}
\begin{tabular}{lcc}
\toprule
& Coefficient on impartial administration & $p$-value \\
\midrule
Pooled OLS (+ year dummies) & $+0.184$ (0.059) & 0.002 \\
Country + Year FE (their spec) & $+0.153$ (0.108) & 0.158 \\[3pt]
\midrule
ICC of impartial administration (unit\,$+$\,year absorbed) & \multicolumn{2}{c}{0.775} \\
Sign test & \multicolumn{2}{c}{Pass (both positive)} \\
Corrected $\hat\lambda_w$ at $\lambda = 0.85$ & \multicolumn{2}{c}{0.33} \\
Debiased FE at $\lambda = 0.85$ & \multicolumn{2}{c}{$+0.458$} \\
Bounds at $\lambda \in [0.85, 0.95]$ & \multicolumn{2}{c}{$[+0.184,\, +0.458]$} \\
Excludes zero? & \multicolumn{2}{c}{YES across all defensible $\lambda$} \\
Verdict & \multicolumn{2}{c}{rescue} \\
\bottomrule
\end{tabular}
\begin{tablenotes}[flushleft]\small
\item \textit{Notes:} 12,048 country-years, 163 countries. DV is five-year-ahead GDP per capita growth. Controls: log GDP per capita, year fixed effects. SEs clustered by country.
\end{tablenotes}
\end{threeparttable}
\end{table}

\begin{figure}[!htbp]
\centering
\begin{tikzpicture}[font=\footnotesize,
  dec/.style={draw, align=center, inner sep=4pt, minimum width=24mm},
  res/.style={draw, rounded corners, align=center, inner sep=4pt, fill=gray!10},
  ->, >={Latex[length=2mm]}]
\node[dec] (d1) at (0,0) {pooled and FE\\same sign?};
\node[dec] (d2) at (0,-1.9) {$|\hat\beta_{FE}| < |\hat\beta_P|$?};
\node[dec] (d3) at (0,-3.8) {reliability\\available?};
\node[dec] (d4) at (0,-5.7) {bounds\\exclude zero?};
\node[res] (r1) at (5,0) {sign flip:\\report FE};
\node[res] (r2) at (5,-1.9) {complement:\\one-sided floor, report FE};
\node[res] (r3) at (5,-3.8) {frontier:\\ME cannot explain shrinkage};
\node[res] (r5) at (5,-5.7) {rescue};
\node[res] (r6) at (0,-7.3) {not identified};
\draw (d1) -- node[above,font=\scriptsize]{no} (r1);
\draw (d1) -- node[left,font=\scriptsize]{yes} (d2);
\draw (d2) -- node[above,font=\scriptsize]{no} (r2);
\draw (d2) -- node[left,font=\scriptsize]{yes} (d3);
\draw (d3) -- node[above,font=\scriptsize]{no} (r3);
\draw (d3) -- node[left,font=\scriptsize]{yes} (d4);
\draw (d4) -- node[above,font=\scriptsize]{yes} (r5);
\draw (d4) -- node[left,font=\scriptsize]{no} (r6);
\end{tikzpicture}
\caption{The diagnostic rule.}
\label{fig:routing}
\end{figure}
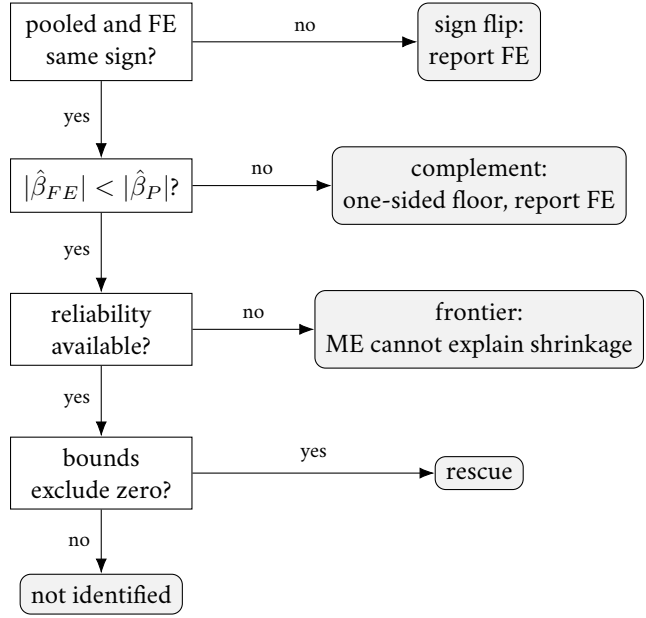

Figure~\ref{fig:routing} depicts the workflow, and Table~\ref{tab:audit} applies it to every application in the larger audit of the literature.

\begin{table}[!htbp]
\centering
\footnotesize
\begin{threeparttable}
\caption{The within-country verdict across applied and famous cases}\label{tab:audit}
\setlength{\tabcolsep}{3pt}
\begin{tabular}{llccclp{3.1cm}}
\toprule
Study/debate & Regressor & Pooled & FE & ICC/floor & Verdict & Substantive interpretation \\
\midrule
CKT bureaucracy   & impartial admin.   & $0.184^{**}$ & $0.153$       & 0.78 & rescue         & bounds (IM $[0.08,1.01]$) exclude zero \\
Haber-Menaldo     & fiscal reliance       & $-0.062$      & $-0.015$       & $0.67^{\dagger}$ & certified      & floor $0.67>r=0.24$; ME ruled out \\
Andersen-Doucette & lag $\Delta$bq        & $-0.054^{*}$  & $-0.181^{***}$ & 0.57 & complement     & FE confirms; reliability floor not identified \\
Brooks-Kurtz      & $\ln$ oil rev pc      & $-1.81^{*}$   & $0.75^{**}$   & 0.69 & sign flip      & estimators disagree; FE reported \\
Democracy-growth  & V-Dem polyarchy       & $0.60^{***}$ & $0.07$        & 0.81 & not identified & IM $[-0.14,0.88]$: null to full effect \\
Democracy-growth  & Polity2               & $0.29^{*}$   & $-0.04$        & 0.64 & sign flip      & second measure flips; FE reported \\
\bottomrule
\end{tabular}
\begin{tablenotes}[flushleft]\footnotesize
\item \textit{Notes:} Two-way fixed effects, SEs clustered by country. Stars mark conventional significance on the pooled and fixed effects coefficients ($^{*}p<0.05$, $^{**}p<0.01$, $^{***}p<0.001$).

\end{tablenotes}
\end{threeparttable}
\end{table}

Table~\ref{tab:audit} shows that the bureaucracy result is the only case in the audit where the correction overturns a fixed effects null. In the other cases, the workflow either leaves the substantive conclusion unresolved, rules out measurement error as the explanation for shrinkage, or sends the result back to fixed effects because the pooled and within unit estimates have opposite signs. To summarize, a fixed effects coefficient by itself does not tell the reader whether shrinkage reflects confounding, attenuation, or a different within unit relationship.

\section{Scope and Sensitivity}\label{sec:scope_alt}

This diagnostic framework applies to typical panel models with continuous, slow moving, mismeasured regressors. Dynamic panels raise additional issues because Nickell bias and measurement error attenuation interact, and generalized methods of moments estimators use both within and between variation \autocite{arellano1991, blundell1998,nickell1981}. The framework also does not carry over directly to nonlinear models, although the same mechanism remains: fixed effects can leave the coefficient identified from noisier variation \autocite{hyslop2001}. This literature reaches a related conclusion about this tradeoff by separating within and between slopes. The diagnostic above adds the measurement error version of that problem \autocite{mundlak1978, belljones2015, kropkokubinec2020}.

Also, the estimated $\lambda_w$ is a marginal diagnostic. In a multivariate regression, the coefficient is scaled by the partial within reliability of the regressor after accounting for the controls. In the application, the partial value differs from the reported $\lambda_w$ by at most about $0.04$, and measurement error in the controls is a separate issue (Appendix~\ref{app:partial}). The i.i.d. formula is also approximate in finite panels. Persistent measurement error raises the true within reliability, while using realized within variance can lower it. In the appendix, I conduct simulations to show that these effects partly offset, and the resulting bounds are conservative (Appendix~\ref{app:ar1_finite_T}).

\subsection{When measurement error is not classical}\label{sec:differential_me}

The propositions above assume classical error. In that model, $u_{it}$ is uncorrelated with the true regressor, the outcome shock, and the outcome itself. That assumption is often unrealistic for variables created by human coders. V-Dem and Polity coders read the same historical record that researchers later use to explain outcomes, including wars, coups, economic crises, mass protests, and bureaucratic quality. If a country is downgraded for institutional quality in the same years that its economic growth deteriorates, then some of the measurement error may be correlated with the outcome. This situation would create differential measurement error rather than classical measurement error.

I treat this issue as a sensitivity problem. Let $\gamma$ be the correlation between the regressor's measurement error and the outcome. The classical model assumes $\gamma = 0$. One can construct differential measurement error bounds by varying $\gamma$ over $[-\gamma_{\max}, \gamma_{\max}]$ together with the reliability range and report the enlarged identified set (Appendix~\ref{sec:classical_me_si}). The useful quantity is the value of $|\gamma|$ at which the substantive conclusion changes. For Cornell-Knutsen-Teorell, the ``rescue'' survives until about $|\gamma| = 0.15$. Beyond that point, one should interpret the case as unidentified. Such a sensitivity analysis shows how much differential error each verdict can tolerate.

Two further complications are handled in the Appendix. Serially correlated measurement error leaves more within variation as signal, so the i.i.d. bounds are conservative and tighten when persistent error is allowed (Appendix~\ref{app:serial_me}). The bounds are also robust to heterogeneous treatment effects. With classical measurement error, they cover the variance weighted estimand (Appendix~\ref{app:hetero}). Finally, tools to conduct sensitivity analyses, such as \textcite{cinelli2020} and \textcite{oster2019}, answer a different question. They ask how strong an omitted confounder would have to be to overturn a result. The bounds here ask how much measurement error changes what fixed effects identify.

\section{Conclusion}

Two-way fixed effects models are the implied standard in comparative politics and international relations. Under classical measurement error, however, fixed effects estimates are attenuated toward zero. For the slow-moving regressors common in this literature, that attenuation can absorb more than half of the coefficient. Substantive conclusions drawn from fixed effects shrinkage often rest on an unexamined measurement assumption: that the regressor is measured without error, or $\lambda = 1$.

The workflow in this paper unmasks this assumption. Researchers should report pooled OLS, fixed effects, the empirical ICC of the key regressor, and the within reliability that scales the attenuation. When pooled OLS and fixed effects share a sign and fixed effects is smaller in magnitude, researchers should report the identified set implied by the defended reliability range. Its width, and whether it excludes zero, should inform the substantive conclusion. When the signs differ, the bounds do not apply and the fixed effects estimate should be reported as a within-unit estimate. When the fixed effects estimate is larger, the case is a complement rather than an attenuation problem. For variables that come with no uncertainty estimates and no second measure, researchers should not invent a reliability number. If measurement error may persist over time, the data cannot separate real stability from stable error. In those cases, researchers should report the within-reliability frontier. The entire workflow is implemented in the R package \texttt{ferobust}. Appendix~\ref{app:ferobust_output} shows example output.

The recent two-way fixed effects literature documents pathologies from heterogeneous effects under binary treatments \autocite{goodmanbacon2021, callaway2021, sunandabraham2021}. The continuous, slow moving regressors studied here carry a different and older pathology: measurement error amplification. The bureaucracy and growth example illustrates the stakes. The fixed effects estimate is consistent with no effect, but the corrected set is also consistent with an effect as large as the cross sectional association. The fixed effects coefficient alone cannot distinguish those conclusions. Reporting the identified set, rather than a lone coefficient, lets readers see whether the within unit verdict is identified.

\section*{Data Availability Statement}

The diagnostic is implemented in the open-source R package \texttt{ferobust} \autocite{ferobust2026}. It is available at \url{https://github.com/asrosenberg/ferobust} and has been submitted to CRAN.

\clearpage
\doublespacing
\printbibliography

\clearpage
\begin{appendices}
\doublespacing
\setcounter{secnumdepth}{2}  % appendices: restore numbered headers

\section{Proofs}\label{app:proofs}

\subsection*{Proof of Proposition~\ref{prop:frontier}}
Because $z$ is the sum of signal and error, and because the signal and error are uncorrelated, the autocorrelation of $z$ is a variance weighted average of the autocorrelations of the two components:
\[
\rho_z(k)
=
\lambda_w \rho_s(k)
+
(1-\lambda_w)\rho_e(k).
\]
The signal autocorrelation is at most one, and the error autocorrelation is at most $\psi_{\max}^{\,k}$. Therefore,
\[
\rho_z(k)
\leq
\lambda_w
+
(1-\lambda_w)\psi_{\max}^{\,k}.
\]
Solving this inequality for $\lambda_w$ gives
\[
\lambda_w
\geq
\frac{\rho_z(k) - \psi_{\max}^{\,k}}{1 - \psi_{\max}^{\,k}}.
\]
This bound holds at every lag $k$, so the sharpest lower bound uses the lag that gives the largest value. Since $\lambda_w$ cannot be negative, the useful lower bound is the maximum of this value and zero. As $\psi_{\max}$ approaches one, the error may be as persistent as the signal, the autocorrelation function no longer separates the two, and the lower bound falls to zero.

\subsection*{Proof of Corollary~\ref{cor:certify}}
The de-attenuated estimate $\hat\beta_{FE}/\lambda_w$ is largest in magnitude at the smallest admissible within-reliability, which the frontier places at $\underline{\lambda}_w(\psi_{\max})$. Its magnitude equals $|\hat\beta_P|$ exactly when $\lambda_w = |\hat\beta_{FE}/\hat\beta_P| = r$. Because the frontier admits only $\lambda_w \ge \underline{\lambda}_w(\psi_{\max})$, a value $\lambda_w \le r$ is admissible if and only if $\underline{\lambda}_w(\psi_{\max}) \le r$. Therefore $\underline{\lambda}_w(\psi_{\max}) > r$ implies $|\hat\beta_{FE}/\lambda_w| < |\hat\beta_P|$ throughout the admissible range.

\section{Partial Identification Bounds}\label{app:bounds}

Under (A1) confounding biases pooled OLS away from zero, (A2) ME attenuates FE toward zero, and (A3) reliability lies in $[\lambda_{\min}, \lambda_{\max}]$. The constraints admit two readings. \emph{Joint reading:} at each $\lambda$, (A2) point-identifies $\beta$ at $\hat\beta_{FE}/\lambda_w(\lambda, \text{ICC})$ in plim and (A1) requires that value to lie at or below $\hat\beta_P$ in magnitude. Both must hold, so only $\lambda$ values satisfying $\hat\beta_{FE}/\lambda_w(\lambda, \text{ICC}) \leq \hat\beta_P$ contribute to the identified set. The strict sharp set is $\mathcal{B}_{\text{strict}} = [\hat\beta_{FE}/\lambda_w(\lambda_{\max}, \text{ICC}),\, \hat\beta_P]$ (when non-empty, i.e., when the breakdown reliability $\lambda^* \leq \lambda_{\max}$). \emph{One-sided reading:} treat (A2) as a one-sided bound (FE attenuates, so $\hat\beta_{FE}/\lambda_w$ bounds $|\beta|$ from below) and (A1) as the other-sided bound from $\hat\beta_P$. Both bounds are valid regardless of which binds at each $\lambda$, and the conservative outer set is the union of the [min, max] intervals: $\mathcal{B} = \bigcup_\lambda [\min(\hat\beta_{FE}/\lambda_w, \hat\beta_P), \max(\hat\beta_{FE}/\lambda_w, \hat\beta_P)]$. The one-sided reading is honest about sampling-error uncertainty in $\hat\beta_P$ (which the joint plim reading conflates with the true plim) and is the bound the \texttt{ferobust} package and Table~\ref{tab:cornell} report. For Cornell-Knutsen-Teorell the breakdown reliability $\lambda^* = 0.962$ exceeds $\lambda_{\max} = 0.95$, so the strict joint set is empty: the de-attenuated estimate is at or above pooled OLS at every defended reliability, and the one-sided reading carries the result, giving $[+0.184, +0.458]$, which excludes zero. The Imbens-Manski CI \autocite{imbens2004} provides correct coverage uniformly over the bound it is built from.

\paragraph{(A1)-relaxed bounds across the applications.}
The half-open relaxed bound $\mathcal{B}_{\text{rel}} = [\hat\beta_{FE}/\lambda_w(\lambda_{\max}, \widehat{\text{ICC}}),\, \infty)$ on the sign of $\hat\beta_{FE}$ asks what survives if (A1) is rejected and pooled OLS no longer plays the role of an upper envelope. For Cornell-Knutsen-Teorell, $\mathcal{B}_{\text{rel}} = [+0.197, +\infty)$, which still excludes zero. The omitted variables that would push pooled OLS away from zero (institutional quality, rule of law, education) are positively correlated with bureaucratic quality and with growth, and reverse causation (growth improving bureaucratic quality) operates in the same direction, so (A1) is defensible; the relaxed bound is reported here as a robustness exercise rather than the operative bound. For Andersen-Doucette no relaxed bound applies: the diagnostic reports fixed effects alone, because the regressor is a difference whose reliability the levels-based Pemstein interval cannot supply (Appendix~\ref{app:lambda_delta}). For Brooks-Kurtz the sign test sends the researcher to FE rather than to bounds, and (A1) does not apply: pooled OLS and FE disagree on sign, the bounds machinery is not the operative tool, and the cross-sectional confounders (regional location, colonial history, ethnic structure) push in too many directions to support a signed-confounding argument either way.

\subsection{Certifying that a null is not measured away: natural resources and authoritarianism}\label{sec:habermenaldo}

The resource curse illustrates the frontier's more common use, which is to confirm that a within-country null reflects a genuine absence of effect rather than attenuation by measurement error. \textcite{haber2011} reappraise the resource curse and conclude that resource wealth does not cause authoritarianism. Their evidence is the disappearance of the cross-national association once country fixed effects are added. The regressor is a constructed point estimate---fiscal reliance on resource revenue---with no measurement model and no parallel measure, so reliability is not identified and Proposition~\ref{prop:frontier} supplies the within-reliability. Replicating their specification on their data, I report the diagnostic in Table~\ref{tab:habermenaldo}.

\begin{table}[!htbp]
\centering\footnotesize
\begin{threeparttable}
\caption{The frontier certifies the Haber-Menaldo null}\label{tab:habermenaldo}
\begin{tabular}{lccccccl}
\toprule
Regressor ($\to$ Polity) & ICC & Pooled & FE & $\rho_z(1)$ & $r$ & $\underline{\lambda}_w(0.7)$ & Verdict \\
\midrule
Fiscal reliance        & 0.46 & $-0.062$ & $-0.015$ & 0.90 & 0.24 & 0.67 & certified\\
Total oil income pc    & 0.57 & $-0.265$ & $-0.097$ & 0.96 & 0.37 & 0.87 & certified \\
\bottomrule
\end{tabular}
\begin{tablenotes}[flushleft]\footnotesize
\item \textit{Notes:} Two-way fixed effects, standard errors clustered by country. Pooled and fixed-effects estimates carry clustered SEs of $0.038$ and $0.032$ (fiscal reliance) and $0.203$ and $0.157$ (oil income); neither fixed-effects coefficient is significant ($p = 0.65$ and $0.54$). $r$ is the shrinkage ratio $|\hat\beta_{FE}/\hat\beta_P|$; $\underline{\lambda}_w(0.7)$ is the frontier floor allowing error persistence up to $\psi_{\max}=0.7$. In both rows $\underline{\lambda}_w(0.7) > r$, so Corollary~\ref{cor:certify} certifies the result.
\end{tablenotes}
\end{threeparttable}
\end{table}

The shrinkage supports a measurement error argument, with the fixed effects coefficient a fraction of the pooled one, and a critic could argue that the within transformation simply mismeasured resource dependence and erased a real curse. The frontier answers this critic's argument. Resource dependence is highly persistent within countries, $\rho_z(1) = 0.90$ for fiscal reliance and $0.96$ for oil income, so even allowing measurement error to be as persistent as $\psi_{\max}=0.7$ the within-reliability stays at or above $0.67$, far above the shrinkage ratios of $0.24$ and $0.37$. By Corollary~\ref{cor:certify} the de-attenuated estimate cannot reach the cross-sectional association. Under this persistence assumption, measurement error alone cannot explain the gap between the cross-sectional and fixed-effects estimates. The diagnostic does not overturn \textcite{haber2011}. It removes the one objection their fixed effects design cannot answer on its own, which is whether measurement error can plausibly explain an absent within-unit effect.

\subsection{The other regimes: complement and sign flip}\label{sec:doucette}

\textcite{andersen2022} argue that bureaucratic quality protects democracies from breakdown, and identify the effect within country. Their preferred specification regresses breakdown onset on the lagged within-country change in bureaucratic quality with country and year fixed effects. The replication yields an estimate of $-0.181$ (SE $0.048$) against the reported $-0.175$. The sign test passes and the fixed effects estimate is \emph{larger} in magnitude than pooled OLS ($-0.181$ versus $-0.054$).

The complement regime would normally add a one-sided floor on $|\beta|$ from the de-attenuated estimate, but here it cannot do so. The regressor is a difference, not a level, and differencing a persistent, noisy series sharply lowers reliability. At the within-country persistence of \texttt{v2clrspct} here ($\hat\phi \approx 0.92$), the Pemstein \emph{levels} interval $\lambda \in [0.85, 0.95]$ implies a differenced reliability of only $\lambda_\Delta \in [0.32, 0.61]$ (Appendix~\ref{app:lambda_delta}). Across nearly the whole range this reliability lies below the change regressor's own ICC of $0.57$ (it rises above only at $\lambda = 0.95$), so the corrected within-reliability turns negative and classical ME is rejected. Thus, I report fixed effects alone.

\textcite{brooks2022} estimate the effect of log oil revenue per capita on Polity2 in 97 countries. Pooled OLS gives $-1.81$, the textbook resource curse; country and year fixed effects give $0.75$, oil booms coinciding with democratization within countries over time. The sign flips, so the diagnostic rule points the researcher to use fixed effects: pooled OLS and the de-attenuated estimate point in opposite directions, and any interval between them would average two contradictory signals. Oil revenue is persistent within countries, so the frontier reads its within-reliability at $\lambda_w \geq 0.75$ under transitory error, which means the within signal is mostly real.

\section{Extended Simulation}\label{app:extended_sim}

I simulate panels varying $N \in \{50, 100, 200\}$ and $T \in \{10, 20, 40\}$ with $\rho \in \{0.3, 0.5, 0.8\}$, ICC $\in \{0.70, 0.85, 0.95\}$, and $\lambda \in \{0.85, 0.90, 0.95\}$, giving 243 cells with 500 replications each.

FE confidence intervals cover $\beta$ in 4.8\% of replications on average against a nominal 95\%, which motivates partial-identification bounds for high-ICC variables.

\subsection*{Grid extensions: unbalanced panels, small-N regional samples, and high-persistence short-T}

The simulations above span balanced panels with $N \geq 50$ and $T \geq 10$. CP and IR researchers sometimes work outside that idealized scenario in short, small-$N$ regional samples (Latin America, post-Soviet states), unbalanced panels from heterogeneous data sources, and dyads or country-years with persistence near a unit root and short observation windows. I run three extension simulations to check that the bounds machinery survives.

\textit{Regime A (unbalanced).} $N = 100$, $T = 20$, $\phi = 0.85$, ICC $\in \{0.70, 0.85\}$, $\lambda = 0.85$, $\rho = 0.5$, 500 replications per cell. Missingness pattern: 30\% missing under MCAR; 30\% missing under selection-on-fixed-effects (units with larger $\alpha_i$ more likely to be missing). The bounds cover the true $\beta$ 100\% of the time in every cell, including the selection-on-FE cells where the unbalanced sample is non-randomly correlated with the unit effect. Imbens-Manski 95\% CI coverage is 100\%.

\textit{Regime B (small $N$).} $N \in \{20, 30, 50\}$, $T \in \{10, 20\}$, $\phi = 0.85$, ICC $\in \{0.70, 0.85\}$, $\lambda = 0.85$, 500 replications per cell. Bounds coverage is at or above 91\% in 10 of 12 cells. The two cells where bounds coverage falls below 90\% are both at the smallest panel size ($N = 20$, $T = 10$): bounds coverage of 85\%--86\%, Imbens-Manski CI coverage of 97\%--99\%. The Imbens-Manski coverage holds at or above the proper value in every cell because the CI widens to absorb the sampling variability of the bounds at small $N$. The implication is conservative: with fewer than $\sim 20$ units in short panels, researchers should report the Imbens-Manski CI rather than the bounds without uncertainty.

\textit{Regime C (high persistence, short $T$).} $N = 100$, $\phi = 0.95$, $T \in \{10, 15, 20\}$, ICC $\in \{0.70, 0.85\}$, $\lambda = 0.85$, 500 replications per cell. This is the scenario that Appendix~\ref{app:ar1_finite_T} highlights as the case where the corrected formula overstates $\lambda_w$. Bounds coverage of the true $\beta$ is 100\% in every cell. The only quantity that degrades in this corner is the sign-test pass rate (77\% at $T = 10$, ICC $= 0.85$, rising to 97\% by $T = 20$): when the within signal is overwhelmed by AR(1) noise in a short panel, pooled OLS and FE disagree on sign more often, and the diagnostic routes more samples to bounds. The bounds themselves still cover.

In short, across the three regimes the bounds machinery does not break in the panels CP and IR researchers actually work with, including unbalanced data and small-$N$ short-$T$ regional samples. The only failure mode is a modest under-coverage of the bare bounds (not the Imbens-Manski CI) when both $N$ and $T$ are at their smallest values simultaneously. The \texttt{ferobust} package highlights this situation and recommends reporting the Imbens-Manski CI alongside the bounds.

\paragraph{DGP.} The DGP for the grid and its extensions is $y_{it} = \alpha_i + \beta x^*_{it} + \varepsilon_{it}$ with $\beta = 0.5$ and $\varepsilon_{it} \sim N(0,1)$. The true regressor follows a unit-specific AR(1) calibrated to the target ICC. Confounding is introduced by $\alpha_i = \rho\mu_i + \sqrt{1-\rho^2}\nu_i$ with $\mu_i$ the unit mean of $x^*$. Measurement error is $u_{it} \sim N(0, \sigma_u^2)$ with $\sigma_u^2$ calibrated to the target $\lambda$. FE is fit via two-way demeaning; pooled OLS uses all variation.

\section{Reliability Estimates and the Literature Audit}\label{app:reliability}

The literature audit shows that many political science variables have high ICCs, which makes them vulnerable to fixed effects attenuation. Whether that attenuation is severe depends on both the empirical ICC and the variable's reliability. Table~\ref{tab:reliability} pairs ICCs from the ICC audit with published reliability intervals and reports the implied within-reliability from Equation~\ref{eq:within_rel}.

\begin{table}[!htbp]
\centering
\footnotesize
\begin{threeparttable}
\caption{Implied Within-Reliability for 20 Common Political-Science Variables, with Published Reliability Intervals. \textit{DI} marks data-inconsistent cells ($\lambda_{\min} \leq \widehat{\text{ICC}}$); see notes.}\label{tab:reliability}
\begin{tabular}{llcccl}
\toprule
 & Variable & ICC & $\lambda$ range & $\lambda_w$ range & Reliability source \\
\midrule
\multicolumn{6}{l}{\textit{Quality of Government Standard}} \\
1 & Corruption Perceptions Index & 0.97 & [0.85, 0.95] & [DI, DI] & TI methodology \\
2 & Trade (\% GDP) & 0.95 & [0.75, 0.90] & [DI, DI] & \textcite{feenstra2015} mirror stats \\
3 & Human Development Index & 0.92 & [0.88, 0.96] & [DI, 0.50] & UNDP methodology \\
4 & V-Dem political corruption (QoG) & 0.84 & [0.87, 0.93] & [0.19, 0.56] & \textcite{pemstein2018} \\
5 & Urban population (\%) & 0.85 & [0.90, 0.96] & [0.33, 0.73] & UN Urbanization Prospects \\
6 & Freedom House status & 0.77 & [0.84, 0.92] & [0.30, 0.65] & FH inter-rater \\
7 & V-Dem libdem (QoG) & 0.75 & [0.91, 0.97] & [0.64, 0.88] & \textcite{pemstein2018} \\
8 & V-Dem polyarchy (QoG) & 0.70 & [0.93, 0.99] & [0.77, 0.97] & \textcite{pemstein2018} \\
9 & Polity2 & 0.64 & [0.84, 0.93] & [0.56, 0.81] & \textcite{treier2008} \\
10 & Population (log) & 0.81 & [0.95, 0.99] & [0.74, 0.95] & UN Population Division \\
11 & Inflation & 0.04 & [0.84, 0.92] & [0.83, 0.92] & WDI documentation \\[4pt]
\multicolumn{6}{l}{\textit{V-Dem Country-Year FullOthers v16}} \\
12 & Mean education & 0.81 & [0.80, 0.90] & [DI, 0.47] & \textcite{boltvanzanden2014} \\
13 & V-Dem rule of law & 0.80 & [0.88, 0.94] & [0.40, 0.70] & \textcite{pemstein2018} \\
14 & V-Dem judicial constraints & 0.79 & [0.88, 0.94] & [0.43, 0.71] & \textcite{pemstein2018} \\
15 & V-Dem political corruption & 0.77 & [0.89, 0.95] & [0.52, 0.78] & \textcite{pemstein2018} \\
16 & V-Dem rigorous public admin & 0.66 & [0.86, 0.92] & [0.59, 0.76] & \textcite{pemstein2018} \\
17 & log GDP per capita & 0.66 & [0.88, 0.96] & [0.65, 0.88] & \textcite{johnson2013} \\
18 & V-Dem libdem & 0.63 & [0.92, 0.98] & [0.78, 0.95] & \textcite{pemstein2018} \\
19 & V-Dem polyarchy & 0.56 & [0.94, 0.99] & [0.86, 0.98] & \textcite{pemstein2018} \\
20 & Population & 0.60 & [0.95, 0.99] & [0.87, 0.97] & UN Population Division \\
\bottomrule
\end{tabular}
\begin{tablenotes}[flushleft]\footnotesize
\item \textit{Notes:} ICCs are computed from the \hyperref[sec:scope]{scope audit}. Reliability intervals come from the cited sources; $\lambda_w$ is computed from Equation~\ref{eq:within_rel}. ``DI'' means the lower reliability bound is inconsistent with classical measurement error because it would imply a latent ICC above 1. Polity2's ICC is for the QoG country-year sample, not the dyadic $\min(\text{Polity2})$ measure used elsewhere.
\end{tablenotes}
\end{threeparttable}
\end{table}

Three patterns emerge. First, several high-ICC variables cannot be reconciled with classical measurement error at the lower edge of their published reliability intervals. The problem is that when the published reliability is less than or equal to the observed ICC, Equation~\ref{eq:within_rel} implies a within-reliability of zero or less. That cannot occur under the classical variance decomposition. For CPI, trade as a share of GDP, HDI, and education, the lower reliability bound is therefore too low. Either the relevant reliability is closer to the upper end of the published range, or the error is not classical.

Second, there is important variation across many of the variables. V-Dem polyarchy retains most of its within-country signal while Polity2 loses a meaningful but not overwhelming share of its signal. Variables that change less within countries, including CINC in Figure~\ref{fig:hidden_cost}, lose much more. The main point is therefore not that all FE estimates collapse, but that attenuation varies predictably with persistence and reliability.

Third, differences between the QoG and full V-Dem versions of the same concepts reflect sample composition. QoG covers a shorter post-1946 country-year panel, while the full V-Dem sample extends back to 1900 and includes more political transitions. Those additional transitions increase genuine within-country variation and therefore raise implied within-reliability.

The Andersen-Doucette application is difficult for the same reason. Its preferred regressor is a difference in bureaucratic quality, and the levels reliability estimate does not identify the reliability of that difference. Once the difference reliability is computed directly, it is too low to support a de-attenuated floor. A levels specification on \texttt{v2clrspct} also fails at the lower edge of the Pemstein interval: in the transition sub-sample, $\widehat{\text{ICC}} = 0.855$, so $\lambda_w = (0.85 - 0.855)/(1 - 0.855) < 0$. I therefore report that application as fixed effects alone rather than as a de-attenuated set.

For V-Dem variables, the posterior uncertainty should be read as a lower bound on measurement error. It captures coder disagreement and bridging uncertainty, but not construct validity problems or systematic coder bias shared across raters. The latter is differential measurement error and is handled separately. For this reason, I use a wider defended reliability range rather than treating the posterior interval as a complete account of measurement error.

The calibrated variables in Figure~\ref{fig:hidden_cost} use reliabilities of $0.95$ for V-Dem polyarchy, $0.85$ for bilateral dyadic log trade, $0.97$ for log GDP per capita, $0.87$ for Polity2, and $0.85$ for log CINC. Their empirical ICCs are $0.64$, $0.50$, $0.92$, $0.73$, and $0.81$, respectively. These cases bracket the broader audit: at central reliabilities, the modal variable retains roughly 50\% to 80\% of its coefficient under one way FE, with substantially worse attenuation under two way FE.

\section{Differenced-Regressor Reliability}\label{app:lambda_delta}

The Andersen-Doucette specification (\hyperref[sec:doucette]{the complement case in the main text}) regresses democratic breakdown on the lagged within-country \emph{change} in bureaucratic quality, not on its level. Pemstein's posterior-based reliability interval for \texttt{v2clrspct}, $\lambda \in [0.85, 0.95]$, is a \emph{levels} reliability and does not transfer to a difference. Differencing a persistent series with classical noise lowers reliability, often sharply, and this section derives by how much.

Let the true regressor follow a stationary AR(1) with persistence $\phi$ and normalize $\operatorname{Var}(x^*) = 1$, so the classical noise variance consistent with levels reliability $\lambda$ is $\sigma_u^2 = (1-\lambda)/\lambda$. The first difference of the signal has variance $\operatorname{Var}(\Delta x^*) = 2(1-\phi)$, and the first difference of i.i.d.\ noise has variance $\operatorname{Var}(\Delta u) = 2\sigma_u^2$. The reliability of the differenced regressor is therefore
\begin{equation}\label{eq:lambda_delta}
\lambda_\Delta = \frac{\operatorname{Var}(\Delta x^*)}{\operatorname{Var}(\Delta x^*) + \operatorname{Var}(\Delta u)}
= \frac{1-\phi}{(1-\phi) + \sigma_u^2}
= \frac{\lambda(1-\phi)}{\lambda(1-\phi) + (1-\lambda)} .
\end{equation}
As $\phi \to 0$ the difference is as reliable as the level ($\lambda_\Delta \to \lambda$); as $\phi \to 1$ the signal difference vanishes while the noise difference does not, and $\lambda_\Delta \to 0$. The more persistent the level, the less a one-period change reveals about the true change.

For \texttt{v2clrspct} in the Andersen-Doucette estimation sample (\texttt{sample3 == 1}, complete cases), a within-country AR(1) regression gives $\hat\phi = 0.92$ (the lag-1 within-country autocorrelation agrees to two digits). Classical noise attenuates observed persistence, so this is if anything a lower bound on the signal persistence, and Equation~\ref{eq:lambda_delta} errs toward \emph{overstating} $\lambda_\Delta$. Evaluated over the Pemstein levels interval, with the change regressor's own empirical $\widehat{\operatorname{ICC}} = 0.57$:
\begin{center}
\begin{tabular}{cccc}
\toprule
levels $\lambda$ & $\lambda_\Delta$ (Eq.~\ref{eq:lambda_delta}) & $\lambda_w = (\lambda_\Delta - \widehat{\operatorname{ICC}})/(1-\widehat{\operatorname{ICC}})$ & data-consistent? \\
\midrule
0.85 & 0.32 & $-0.58$ & no \\
0.90 & 0.43 & $-0.33$ & no \\
0.93 & 0.53 & $-0.10$ & no \\
0.95 & 0.61 & $+0.10$ & yes \\
\bottomrule
\end{tabular}
\end{center}
A positive corrected within-reliability requires $\lambda_\Delta > \widehat{\operatorname{ICC}} = 0.57$, which holds only at the very top of the Pemstein interval, near $\lambda = 0.95$. Across most of the defended levels range $\lambda_\Delta$ falls below the change regressor's ICC, the corrected $\lambda_w$ is negative, and the classical measurement-error model is rejected: there is no admissible within-reliability with which to de-attenuate the fixed-effects estimate. The one-sided floor the complement regime would supply therefore requires a change-regressor reliability the levels-based Pemstein interval cannot provide, and the application is reported as fixed effects alone. Even at the favorable upper edge the implied de-attenuated coefficient is implausibly large for a breakdown-probability model ($|\hat\beta_{FE}|/\lambda_w \approx 1.8$ at $\lambda = 0.95$, and larger still below it), a further sign that a differenced-classical-ME reading is strained. Reproduced by \texttt{R/doucette\_change\_spec\_diagnostic.R}.

\section{Correlated Random Effects Under Measurement Error}\label{app:cre}

The correlated random effects (Mundlak) estimator adds unit means of the regressors to a random-effects specification, absorbing time-invariant confounding into $\bar{x}_i$. Its within coefficient is numerically the fixed-effects estimate on $x_{it}$ and therefore inherits the same attenuation, converging to $\beta\lambda_w$ by Equation~\ref{eq:within_rel}. For a question about within-unit effects, CRE is thus no better than fixed effects under measurement error, which is why the diagnostic works with the fixed-effects coefficient directly rather than substituting CRE.

\section{Heterogeneous Treatment Effects and ME}\label{app:hetero}

With heterogeneous effects, the FE estimand is not the unweighted mean effect
but a variance-weighted average,
\[
  \beta_{VW} = \frac{\sum_i w_i \beta_i}{\sum_i w_i},
\]
where $w_i$ is unit $i$'s within-unit variance in the regressor
\autocite{imaikim2019, dechaisemartin2020}. Under the classical
measurement-error structure considered here, the within-reliability factor
$\lambda_w$ attenuates the FE coefficient, so the probability limit of the
observed estimate is $\beta_{VW}\lambda_w$. The bounds still apply; their
target is simply $\beta_{VW}$ rather than the unweighted mean $\bar\beta$.

I assess this claim with 14{,}400 simulated panels. The simulations vary the
scale of treatment-effect heterogeneity,
$\sigma_\beta \in \{0,\, 0.25\bar\beta,\, 0.5\bar\beta,\, \bar\beta\}$;
the empirical ICC, $\widehat{\mathrm{ICC}} \in \{0.70, 0.85, 0.95\}$; and
reliability, $\lambda \in \{0.85, 0.90, 0.95\}$. All simulations set $N=100$,
$T=20$, and $\bar\beta = 0.5$, with classical measurement error at reliability
$\lambda$. For each panel, I construct the bounds with the corrected formula
over the assumed reliability interval $[\lambda - 0.05,\, \lambda + 0.05]$ and
check whether the interval covers $\beta_{VW}$.

Coverage of $\beta_{VW}$ remains stable across heterogeneity levels: 97.1\%
when $\sigma_\beta = 0$, 96.8\% at $0.25\bar\beta$, 97.2\% at $0.5\bar\beta$,
and 97.1\% at $\bar\beta$. Coverage of the unweighted mean $\bar\beta$ falls
under strong heterogeneity: 97.1\%, 96.6\%, 95.7\%, and 93.9\%, respectively.
This decline is expected because $\bar\beta$ is not the FE estimand when
treatment effects covary with the regressor's within-unit variance. The
sign-test rule passes in 99.7\% to 99.9\% of simulations at every
heterogeneity level. The diagnostic is therefore robust to treatment-effect
heterogeneity in these simulations, provided one interprets its target as
$\beta_{VW}$ rather than $\bar\beta$.

\textit{Small-$T$ check.} Re-running the same simulation at $T \in \{10, 15\}$ (43,200 additional panels) shows the picture worsens in the expected way as the within-unit sample shrinks. At $T = 20$, bounds coverage of $\beta_{VW}$ averages 96.6\%--97.1\% across heterogeneity levels; at $T = 15$, 94.4\%--94.9\%; at $T = 10$, 89.6\%--90.4\%. The drop from $T = 20$ to $T = 10$ exceeds the 5-percentage-point threshold I treat as the scope condition. The sign-test rule, by contrast, holds up: pass rates remain at 98.8\%--99.2\% at $T = 10$ and 99.4\%--99.8\% at $T = 15$, with the only meaningful drop at $\widehat{\text{ICC}} = 0.95$ and $T = 10$ (96.8\%). Two practical implications. First, the routing rule is robust to small $T$: the recommendation between fixed effects and bounds does not change in the panels CP and IR researchers work with. Second, the bounds at $T < 20$ are under-covering the variance-weighted FE estimand by 5--7 percentage points, so researchers with short panels should report the Imbens-Manski CI rather than the bare bounds; the \texttt{ferobust} package issues a small-$T$ warning when $T < 20$.

\section{Marginal versus Partial Within-Reliability}\label{app:partial}

The corrected within-reliability $\lambda_w$ is computed from the key regressor's marginal ICC. In a multivariate fixed-effects regression the coefficient on the key regressor attenuates by its \emph{partial} within-reliability, the reliability of the part of the regressor orthogonal to the controls after two-way demeaning. By the Frisch-Waugh-Lovell theorem the relevant signal is the within-unit variation in the key regressor not explained by the controls, while classical measurement error, being orthogonal to the controls, survives the partialling undiminished. Writing $R^2_{x\mid W}$ for the within-unit $R^2$ of the observed key regressor on the observed controls, the partial within-reliability is
\begin{equation}
\lambda_w^{\text{part}} = \frac{\lambda_w(1 - R^{2}_{*})}{\lambda_w(1 - R^{2}_{*}) + (1 - \lambda_w)}, \qquad R^{2}_{*} = \frac{R^2_{x\mid W}}{\lambda_w},
\end{equation}
where $R^{2}_{*}$ is the implied within-$R^2$ for the latent regressor, since the observed $R^2$ is itself attenuated by the noise in the key regressor. The gap between marginal and partial reliability is governed by $R^2_{x\mid W}$: when the controls explain little within-unit variation in the key regressor, the two coincide.

In the worked applications $R^2_{x\mid W}$ is small. Regressing \texttt{v2clrspct} on log GDP per capita with country and year fixed effects gives $R^2_{x\mid W} = 0.058$; regressing V-Dem polyarchy on log population and average education gives $0.008$. The implied partial reliabilities barely move:
\begin{center}
\begin{tabular}{lcc}
\toprule
Application & marginal $\lambda_w$ & partial $\lambda_w^{\text{part}}$ \\
\midrule
Cornell-Knutsen-Teorell ($\lambda = 0.85$) & 0.33 & 0.29 \\
Cornell-Knutsen-Teorell ($\lambda = 0.95$) & 0.78 & 0.76 \\
Democracy-growth ($\lambda = 0.93$) & 0.63 & 0.63 \\
Democracy-growth ($\lambda = 0.97$) & 0.84 & 0.84 \\
\bottomrule
\end{tabular}
\end{center}
The partial value lies slightly below the marginal one, so using $\lambda_w$ as reported implies marginally less attenuation than the partial calculation and is mildly conservative for the bounds; no verdict changes. Measurement error in the controls is a distinct channel that $\lambda_w$ does
not capture: error in a control leaks into the coefficient on the key
regressor, with a sign that depends on the within-unit covariance structure.
When a mismeasured control covaries strongly with the key regressor within
units, the analyst should compute the partial reliability directly and treat
control error as a separate source of bias.

\section{Differential ME Sensitivity}\label{sec:classical_me_si}

This appendix derives the machinery; the per-application results are reported in the main text (\hyperref[sec:differential_me]{When measurement error is not classical}). Sensitivity exercises at illustrative parameter values (ICC $= 0.85$, $\lambda = 0.87$, $\rho = 0.5$): I correlate the measurement error with either $x^*$ or $\varepsilon$ at strength $\gamma$ and sweep $\gamma \in [0, 0.5]$. When noise correlates with $x^*$, the RE advantage holds across the full grid. When noise correlates with $\varepsilon$, the RE advantage narrows at $\gamma \approx 0.10$. Operationally, a researcher who cannot commit to classical ME sweeps $\gamma \in [-\gamma_{\max}, \gamma_{\max}]$, where $\gamma$ is the correlation between the regressor's measurement error and the outcome, jointly with the reliability range and reports the enlarged identified set. The thresholds quoted in the main text are the $|\gamma|$ at which zero enters each enlarged set: $|\gamma| \approx 0.15$ for the bureaucracy-growth rescue and $|\gamma| \le 0.30$ for the oil-rents sign flip (robust because the pooled and FE coefficients disagree by orders of magnitude).

\section{Serially-Correlated Measurement Error}\label{app:serial_me}

The main derivation assumes the noise $u_{it}$ is uncorrelated across years, so that the entire noise variance lands in the within-unit dimension. For coder-based variables, the error is likely persistent because a country coded overly generously in one year tends to be coded generously in the next, so $u_{it}$ carries a unit-specific component $\bar{u}_i$. Let $\pi$ denote the error's own ICC, which again is the share of its variance that is persistent (between-unit) rather than transitory (within-unit).

\textbf{Derivation.} Normalize $\text{Var}(x^*) = 1$, so $\sigma_u^2 = (1-\lambda)/\lambda$. Split the error as $u_{it} = \bar{u}_i + \tilde{u}_{it}$ with $\text{Var}(\bar{u}_i) = \pi\sigma_u^2$ and $\text{Var}(\tilde{u}_{it}) = (1-\pi)\sigma_u^2$. The persistent component adds to the between variance, so the observed ICC becomes
\[
\widehat{\text{ICC}} = \frac{\text{ICC}^* + \pi\sigma_u^2}{1 + \sigma_u^2} = \lambda\,\text{ICC}^* + \pi(1-\lambda),
\]
which recovers the i.i.d. identity $\widehat{\text{ICC}} = \lambda\,\text{ICC}^*$ at $\pi = 0$. The within-reliability is the within signal over within total, $\lambda_w = (1-\text{ICC}^*)/[(1-\text{ICC}^*) + (1-\pi)\sigma_u^2]$. Substituting $\text{ICC}^* = (\widehat{\text{ICC}} - \pi(1-\lambda))/\lambda$ and simplifying, the denominator collapses to $1 - \widehat{\text{ICC}}$ and the $\pi$-aware within-reliability is
\begin{equation}\label{eq:serial_lw}
\lambda_w(\pi) = \frac{\lambda - \widehat{\text{ICC}} + \pi(1-\lambda)}{1 - \widehat{\text{ICC}}} = \lambda_w(0) + \frac{\pi(1-\lambda)}{1 - \widehat{\text{ICC}}},
\end{equation}
which recovers the i.i.d. formula of Equation~\ref{eq:within_rel} at $\pi = 0$.
Since $\pi \geq 0$ and $\lambda < 1$, persistent error always \textit{raises} $\lambda_w$. The i.i.d. formula is a lower bound on within-reliability and hence an upper bound on FE attenuation. When $\pi = 1$, the error is purely between-unit, so demeaning removes all of the error, and $\lambda_w \to 1$.

\textbf{Monte Carlo validation.} I simulate a panel ($N = 100$, $T = 20$, $\beta = 0.5$, $\text{ICC}^* = 0.85$, $\lambda = 0.90$, $\rho = 0.5$, 400 replications per cell) with the error split into persistent and transitory pieces, and I vary the error's ICC over $\pi \in \{0, 0.2, 0.4, 0.6, 0.8\}$. Table~\ref{tab:serial_me} reports the realized within-reliability (computed from the simulated $x^*$ and $x$), the naive i.i.d. formula evaluated at the observed ICC, and the $\pi$-aware formula of Equation~\ref{eq:serial_lw}. The $\pi$-aware formula tracks the realized value to within about 0.03 across the range (the residual gap is the finite-$T$ AR(1) effect of Appendix~\ref{app:ar1_finite_T}), while the i.i.d. formula falls to 0.20 when the realized value is 0.85.

\begin{table}[!htbp]
\centering
\footnotesize
\begin{threeparttable}
\caption{Within-reliability under serially-correlated measurement error}\label{tab:serial_me}
\begin{tabular}{ccccc}
\toprule
$\pi$ (error ICC) & Observed $\widehat{\text{ICC}}$ & Realized $\lambda_w$ & i.i.d.\ formula & $\pi$-aware formula~\eqref{eq:serial_lw} \\
\midrule
0.0 & 0.799 & 0.538 & 0.503 & 0.503 \\
0.2 & 0.817 & 0.591 & 0.453 & 0.562 \\
0.4 & 0.835 & 0.659 & 0.395 & 0.637 \\
0.6 & 0.857 & 0.745 & 0.303 & 0.721 \\
0.8 & 0.876 & 0.853 & 0.195 & 0.839 \\
\bottomrule
\end{tabular}
\begin{tablenotes}[flushleft]\footnotesize
\item \textit{Notes:} $N = 100$, $T = 20$, $\beta = 0.5$, $\text{ICC}^* = 0.85$, $\lambda = 0.90$, $\rho = 0.5$, 400 replications per cell. ``Realized $\lambda_w$'' is the ratio of within-demeaned signal variance to within-demeaned observed variance in the simulated data.
\end{tablenotes}
\end{threeparttable}
\end{table}

\textbf{Application check.} For the one application where the diagnostic routes to bounds (Cornell-Knutsen-Teorell), I recompute the identified set across $\pi \in [0, 1]$ using $\lambda_w(\pi)$ in place of the i.i.d. within-reliability, holding the published diagnostic inputs ($\hat\beta_{FE}$, $\hat\beta_P$, $\widehat{\text{ICC}}$, reliability range) fixed. Zero remains excluded across the entire range. The set tightens from $[0.184, 0.458]$ at $\pi = 0$ to $[0.153, 0.184]$ at $\pi = 1$ (collapsing toward the raw FE and pooled estimates). Because the bound is widest at $\pi = 0$, the i.i.d. analysis reported in the main text is conservative, and admitting serial correlation in the measurement error only narrows the set without moving it across zero. The sign-flip (Brooks-Kurtz) and complement (Andersen-Doucette) cases above report the fixed effects estimate rather than a set, so this test does not bear on them. For the frontier case (Haber-Menaldo), persistent error only \emph{raises} the within-reliability floor of Proposition~\ref{prop:frontier}.

\section{Within-Reliability Frontier: Validation}\label{app:frontier_val}

This section confirms two claims. First, single-series estimates of overall reliability fail when measurement error is persistent. Second, the frontier in Proposition~\ref{prop:frontier} provides a conservative bound on $\lambda_w$ when the assumed persistence ceiling $\psi_{\max}$ weakly exceeds the true error persistence. If the ceiling is set too low, the bound is no longer guaranteed to be conservative.

The simulations generate the observed regressor as a unit mean plus an AR(1) within-unit signal and measurement error. I vary $N$, $T$, latent ICC, signal persistence, reliability, a common time trend, and panel missingness across five error regimes: i.i.d., weakly persistent, moderately persistent, signal-like, and unit-bias error.

\paragraph{The single-series reliability estimator fails under persistent error.} The AR(1)-plus-i.i.d.-noise estimator recovers overall reliability almost exactly when measurement error is serially uncorrelated: the mean bias is $0.003$, and every replication falls within $0.05$ of the truth. Once the error is persistent, however, the estimator systematically overstates reliability. Mean bias rises to $0.074$ under weak persistence, $0.099$ under moderate persistence, and $0.097$ under signal-like error, with estimates falling within $0.05$ of the truth in only $33\%$, $16\%$, and $18\%$ of replications, respectively. The bias is upward in every persistent-error regime, which is precisely the direction that understates FE attenuation. Two-root extensions that jointly estimate signal and error persistence perform no better: they return valid estimates in only $36\%$ to $53\%$ of the difficult cells and do not reduce bias when they do converge.

\paragraph{The frontier covers conservatively under a defensible $\psi_{\max}$.} Table~\ref{tab:frontier_val} reports the plug-in frontier on the cells where $\psi_{\max}$ is at least the true error persistence. Coverage of the true $\lambda_w$ is $1.00$ in every regime. The bound is loose, as a worst-case bound should be, and it loosens as the error is allowed to be more persistent: the mean lower bound falls from $0.108$ under i.i.d.\ error to near zero when the error may be as persistent as the signal. The bound tightens with panel length, from a mean of about $0.03$ at $T = 10$ to about $0.18$ at $T = 40$, coverage holding at $1.00$ throughout.

\begin{table}[!htbp]
\centering
\footnotesize
\begin{threeparttable}
\caption{Plug-in within-reliability frontier under a defensible persistence ceiling}\label{tab:frontier_val}
\begin{tabular}{lcccc}
\toprule
Error regime & $\psi_{\max}$ used & True $\lambda_w$ & Mean lower bound & Coverage \\
\midrule
i.i.d. error            & 0.0 & 0.483 & 0.108 & 1.00 \\
Weak persistent          & 0.3 & 0.557 & 0.097 & 1.00 \\
Moderate persistent      & 0.7 & 0.592 & 0.020 & 1.00 \\
Signal-like              & 0.9 & 0.673 & 0.000 & 1.00 \\
Unit-bias                & 0.0 & 0.678 & 0.174 & 1.00 \\
\bottomrule
\end{tabular}
\begin{tablenotes}[flushleft]\footnotesize
\item \textit{Notes:} Plug-in sample-autocorrelation frontier, 240 sampled DGP cells, 30 replications each. Rows condition on $\psi_{\max}$ being at least the true error persistence. ``Coverage'' is the share of replications in which the lower bound does not exceed the true $\lambda_w$.
\end{tablenotes}
\end{threeparttable}
\end{table}

\paragraph{The frontier requires a credible assumption about  persistence.} The frontier is conservative only if the researcher assumes measurement error is at least as persistent as it is in the true data generating process. When the assumed ceiling $\psi_{\max}$ is too low, the lower bound can be too high. In the plug-in simulation, weakly persistent error ($\psi_e = 0.3$) covered the true value in only $70\%$ of replications when evaluated under the false assumption $\psi_{\max}=0$. Moderate error ($\psi_e = 0.7$) covered the truth in $95\%$ of replications when evaluated with $\psi_{\max}=0.5$, and signal-like error ($\psi_e = 0.9$) covered the truth in $96\%$ when evaluated with $\psi_{\max}=0.7$. In all three cases, setting $\psi_{\max}$ equal to the true error persistence allowed the technique to recover the truth. The implication is straightforward: the method is conservative when the researcher makes a credible allowance for persistent measurement error, but not when the researcher assumes it away.
\paragraph{Bootstrap variant.} I also consider a finite sample version that takes the fifth percentile of the frontier across unit-bootstrap resamples. This version covers in $100\%$ of replications, but it is too conservative to use as the default because it often collapses to a zero lower bound. I therefore report it as a robustness check rather than as the operative bound.
\section{AR(1) Finite-T Calibration}\label{app:ar1_finite_T}
The corrected within-reliability formula is derived for the population variance decomposition. In short panels, however, highly persistent regressors may not realize their full within-unit variance, which can compromise inferences. When the latent regressor follows an AR(1) process with persistence $\phi$, this finite-$T$ problem can make the formula overstate the amount of usable within-unit signal. The problem is largest when persistence is high and the panel is short.
To quantify this gap, I simulate panels by varying $\phi \in \{0, 0.30, 0.60, 0.85, 0.95\}$, $T \in \{10, 20, 40, 80\}$, and $\widehat{\text{ICC}} \in \{0.70, 0.85, 0.95\}$, while holding fixed reliability $\lambda=0.85$ and $N=200$. In each cell, I compare the realized within-reliability with the value provided by the corrected formula. When $\phi=0$, the formula matches the realized value to within 0.01 across panel lengths. When persistence is high and panels are short, the gap is much larger. At $\phi=0.95$ and $T=20$, for example, the formula predicts $\lambda_w \approx 0.50$, while the realized value is closer to $0.20$. The gap shrinks as $T$ grows. If $T=80$, then the two values only differ by 0.03.
The practical implication of this simulation is that the corrected formula can understate attenuation for highly persistent regressors in short panels. This does not invalidate the bounds. For a positive coefficient, overstating $\lambda_w$ makes the corrected FE estimate smaller, which lowers the de-attenuated endpoint and widens the identified set. Researchers working with highly persistent regressors in panels with $T<40$ should therefore treat the resulting bounds as conservative.
A further simulation confirms that the bounds are conservative. I simulate a scenario where the calibration gap is largest: $\phi \in \{0.85, 0.95\}$, $T \in \{10, 20, 40\}$, and $\widehat{\text{ICC}} \in \{0.70, 0.85\}$, with 1{,}000 panels per cell, $N=200$, $\lambda=0.85$, and $\beta=1$. For each panel, I construct the bounds $B = [\hat\beta_{FE}/\hat\lambda_w,\, \hat\beta_P]$ using the corrected formula evaluated at $\widehat{\text{ICC}}$, and then consider whether the true $\beta$ falls inside both $B$ and the Imbens-Manski 95\% confidence interval.

The bounds cover the true $\beta$ in 99.9\%--100\% of replications in every cell of the grid, including the worst case ($\phi = 0.95$, $T = 10$, $\widehat{\text{ICC}} = 0.85$, where the formula most overstates $\lambda_w$). Imbens-Manski 95\% CIs cover at 100\% in every cell. Because the formula treats realized within variance as if it were population variance, the implied $\hat\lambda_w$ is too high. The corrected FE estimate $\hat\beta_{FE}/\hat\lambda_w$ is therefore too small in magnitude, the lower edge of $B$ is conservative, and the bounds end up wider than necessary. The cost is width, not coverage. When a nuisance parameter is uncertain, partial-identification bounds should err wide, which is what they do here.

\section{Operating Characteristics of the Diagnostic}\label{app:diag_accuracy}

The diagnostic's rule is built from sample quantities (the sign comparison, $\widehat{\text{ICC}}$, and the pooled-FE shrinkage ratio) and inherits their sampling variability. A simulation across 500 panels in each of 64 grid cells records the MSE of the diagnostic's recommended estimator against an oracle baseline. At $\widehat{\text{ICC}} \leq 0.90$ the median ratio of diagnostic-MSE to always-FE MSE is 0.69, and the diagnostic beats the fixed-effects default in 88\% of cells. At $\widehat{\text{ICC}} = 0.95$ the rule recommends bounds in 98.7\% of samples. The rule fails at $\lambda \geq 0.95$ combined with strong confounding ($\rho \geq 0.5$). A researcher facing this situation should rely on a reliability argument over the diagnostic. The \texttt{ferobust} package highlights any cell where the assumed $\lambda$ exceeds 0.95 and prompts a confounding-severity check.

\section{Example \texttt{ferobust} Output}\label{app:ferobust_output}

Applied to the bureaucracy-growth replication of \textcite{cornell2020}, the single function returns the sign-test verdict, the empirical ICC, the corrected within-reliability at each defended reliability, the partial-identification bounds, the Imbens-Manski confidence interval, and the breakdown reliability:

\begin{verbatim}
> ferobust_lm(gdp_growth_F5 ~ v2clrspct + s_mil_loggdp | country_id + year,
+             data = ckt, reliability = c(low = 0.85, mid = 0.90, high = 0.95))

ferobust diagnostic
--------------------------------------------------
Pooled:       0.1839    FE:       0.1532
Shrinkage:   16.7%
Same sign:  yes
ICC (unit): 0.736    ICC (unit+year absorbed): 0.775  [bounds use this two-way ICC]
--------------------------------------------------
Within-reliability:
  at lambda = 0.85:  lambda_w = 0.334
  at lambda = 0.95:  lambda_w = 0.778
Breakdown reliability (lambda*): 0.962
--------------------------------------------------
Identified set:
  at lambda = 0.85:  [0.1839, 0.4582]
  at lambda = 0.95:  [0.1839, 0.1969]
Imbens-Manski 95% CI: [0.0833, 1.0094]
Excludes zero:       yes
\end{verbatim}

\end{appendices}

\typeout{get arXiv to do 4 passes: Label(s) may have changed. Rerun}
\end{document}